**Title page**

**Title**: Rigidity controls human desmoplastic matrix anisotropy to enable pancreatic cancer invasion via extracellular signal-regulated kinase 2

**Short title**: Substrate softening normalizes stromal matrices

**Authors**: R. Malik[1, 2], T. Luong[1], X. Cao[3], B. Han[4], N. Shah[1], L.Han[4], V. B. Shenoy[3], P. I. Lelkes[2*] and E. Cukierman[1*]

**Author Affiliations:**

[1]Cancer Biology Program, Fox Chase Cancer Center

[2]Dept. Bioengineering, Temple University

[3]Materials Science and Engineering, University of Pennsylvania

[4] School of Biomedical Engineering, Science and Health Systems, Drexel University

**Corresponding Authors:**

* Edna Cukierman - Edna.Cukierman@fccc.edu (ORCID 0000-0002-1452-9576)

* Peter I. Lelkes - pilelkes@temple.edu - Laura H. Carnell Professor of Bioengineering at Temple University's College of Engineering (ORCID 0000-0003-4954-3498)

**Keywords:** cancer associated fibroblasts, cell-derived extracellular matrix, desmoplasia, fiber alignment, matrix-induced responses, pancreatic cancer, underlying substrate stiffness

**Abstract**

Pancreatic ductal adenocarcinoma (PDAC) encompasses a fibrous-like tumor-associated stroma, desmoplasia, produced by cancer-associated fibroblasts (CAFs). As desmoplasia plays dual pro- and anti-tumor roles, efforts are being made to induce its anti-tumor functions for therapeutic gain. We conducted biomechanical manipulations, using variations of pathological and physiological substrates *in vitro,* to culture patient-harvested CAFs and generate desmoplastic CAF-derived extracellular matrices (CDMs) that can restrict PDAC growth and invasion. We posited that extrinsic modulation of the environment, via substrate rigidity, influences CAF's cell-intrinsic forces affecting CDM production. Substrates used were polyacrylamide gels of physiological (~1.5 kPa) or pathological (~7.5 kPa) stiffnesses. Results showed that physiological substrates influenced CAFs to generate CDMs similar to normal/control fibroblasts. We found CDMs to be softer than the corresponding underlying substrates, and CDM fiber anisotropy (i.e., alignment) to be biphasic and informed via substrate-imparted morphological CAF aspect ratios. Further, the biphasic nature of CDM fiber anisotropy was mathematically modeled and postulated a correlation between CAF aspect ratios and CDM alignment; regulated by extrinsic and intrinsic forces to conserve minimal free energy. Biomechanical manipulation of CDMs, generated on physiological stiffness substrates, lead to reduction in nuclear translocation of pERK1/2 in KRAS-driven PDAC. Specifically, ERK2 was essential for CDM-regulated tumor cell invasion. *In vitro* findings were validated *in vivo,* where nuclear pERK1/2 was found significantly higher in human PDAC samples compared to matching normal pancreata. The study suggests that altering underlying substrates enables CAFs to remodel CDMs and restricts KRAS-driven PDAC invasion in an ERK2 dependent manner.

**Significance**

The pancreatic ductal adenocarcinoma (PDAC) 5-year survival rate is ~8%; by 2020, PDAC will become the second most lethal cancer in the US. PDAC includes a fibrous-like stroma, desmoplasia, encompassing most of the tumor mass, which is produced by cancer-associated fibroblasts (CAFs) and their cell-derived extracellular matrices (CDMs). Since elimination of CAFs has proven to be detrimental to patients, CDM reprogramming, as opposed to desmoplasia ablation, is therapeutically desirable. Data from this study suggest that the architecture of CDMs, and not necessarily matrix stiffness, could be manipulated to render a tumor-suppressive microenvironment. Hence, treatments that could reprogram desmoplasia to become tumor-suppressive/restrictive or that could target tumoral ERK2 might provide new means for treating PDAC patients in the future.

**\body**

## Introduction

Fibroblastic stromal extracellular matrices (ECMs) modulate essential cellular behaviors such as differentiation, migration, proliferation, and survival [1]. In epithelial cancers such as pancreatic ductal adenocarcinoma (PDAC), loss of the homeostatic equilibrium of normal stroma induces mechanical and biochemical changes, resulting in dynamic activation of fibroblastic pancreatic stellate cells (PSC). In turn, the resultant activated cancer-associated fibroblasts (CAFs) remodel and deposit CAF-derived ECMs (CDMs), generating a dynamically altered fibrotic tumor microenvironment known as desmoplasia [2, 3]. Biomechanical characteristics of desmoplasia, such as ECM fiber anisotropy (i.e., parallel organized CDM fibers), have been correlated with poor cancer survival in numerous epithelial cancers, including PDAC [4-6], yet the mechanisms by which desmoplastic stroma promotes tumor progression remain unclear.

Along with stromal alterations, over 90% of all PDACs encompass KRAS mutations [7-9], which are evident during early phases of PDAC manifestation [10] and essential for both tumor initiation and progression [11, 12]. Despite the high frequency of activating KRAS mutations (e.g., $KRAS^{G12D}$), PDAC development and progression depends on chronic KRAS stimulation, including of the mutant allele, by the tumor-associated desmoplastic stroma [13-15]. Notwithstanding the recognition of the key role of KRAS in the development/progression of PDAC, therapies targeting KRAS signaling have failed to show clinical efficacy [16, 17]. As an alternative approach, targeting the stromal reciprocity that maintains KRAS active [13] may assist in treating this disease.

While CAFs are responsible for producing and remodeling CDMs and maintaining many of the tumorigenic aspects of desmoplasia [18-20], ablation of CAFs is detrimental to patients [21]. Therefore, CAF reprogramming, i.e. harnessing the innate, tumor-restrictive properties of the "normal" microenvironment, as opposed to stromal ablation, is an attractive therapeutic approach [22-26]. Many studies have suggested that pancreatic tumor stiffness is significantly greater than the physiological pancreas [27]. Tissue stiffness is a major factor that regulates naïve-to-CAF activation as well as CAFs' ability to remodel desmoplastic ECMs [28]. The contractility of adherent cells (e.g., myofibroblastic CAFs), in concert with the extracellular physical properties of the substrate, constitute, respectively, the intrinsic and extrinsic forces needed to regulate tissue architecture (e.g., ECM isotropy) [25, 28-31]. Nonetheless, the biomechanical mechanisms that enable a stromal ECM production with tumor-restrictive, as opposed to tumor-permissive, capabilities remain unclear.

This study tests the hypothesis that manipulations altering ECM architecture or intracellular CAF myofibroblastic contractility could result in a "normalized" tumor-restrictive microenvironment. In testing our hypothesis, we first investigated whether modulating substrate rigidity affects the ability of CAFs to adjust CDM fiber anisotropy and/or rigidity. We further questioned whether CDMs could be modified to restrict, rather than promote, KRAS-driven tumorigenicity of PDAC cells. While our previous studies presented a mathematical model to explain the correlation between stiffness, cell polarization and matrix alignment [32], we now modified this model to explain how physiological substrate stiffness triggers isotropic normalization of CDMs. Our results suggest that physiological stiffness can affect CAFs to produce CDMs that restrict tumor cell growth and invasion by preventing nuclear localization of activated extracellular signal-regulated kinase (ERK) in an ERK2 dependent manner.

**Results**

To assess stiffness-modulated CAF morphology, patient-harvested CAFs [33] were cultured on collagen-I conjugated polyacrylamide gels of pancreatic physiological (physio-gel; ~1.5 kPa) or pathological (patho-gel; ~7 kPa) stiffnesses [27] using glass coverslips as a rigid-substrate control. As seen in Figure 1, the cell aspect ratio (i.e., length over width) of CAFs indicated a biphasic distribution, with a peak of about 3 times length over width for CAFs cultured on patho-gels, compared to CAF aspect ratios of 1.8 and 2.3, respectively, measured on physio-gels and glass coverslips. A similar, but discrete, trend was observed when control (i.e., inactive) fibroblasts were used (Supplemental Figure 1A). These results indicated that physiologically soft substrates can limit the aspect ratio of CAFs rendering round, as opposed to spindled, morphologies.

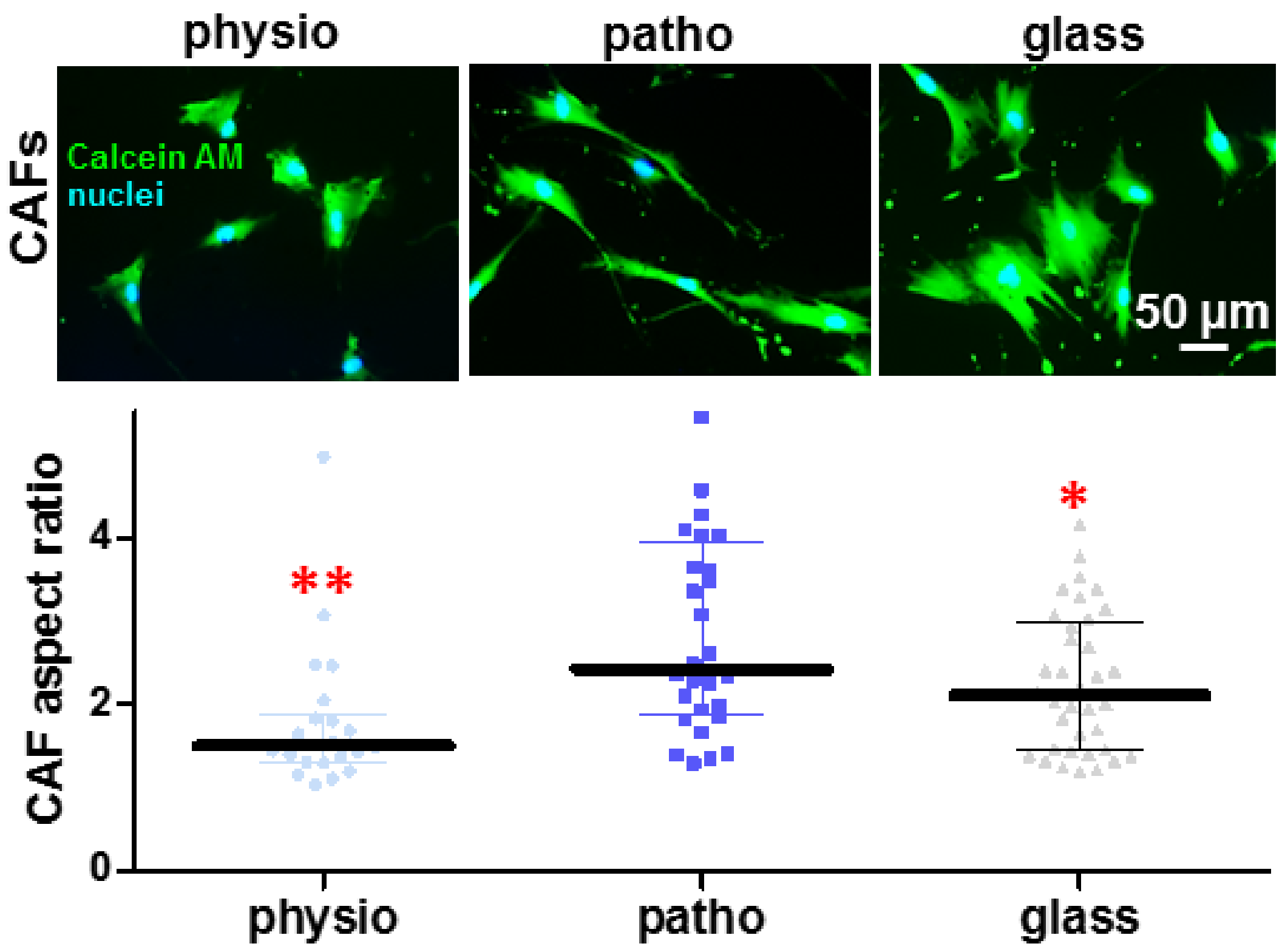

**Figure 1: Changes in substrate stiffness dictate a biphasic distribution of CAF aspect ratios**. Images correspond to epi-fluorescent microscopy acquired CAFs labeled to depict cell morphology (CalceinAM; green) and nuclei (blue) cultured onto physio (~1.5 kPa), patho (~7 kPa) gels or glass coverslips. Graph includes measured CAF aspect ratios (length/breadth) calculated using MetaMorph software (below). Data is presented as median ± interquartile range. Asterisks correspond to * $p<0.05$ and ** $p<0.01$, compared to ratios attained by CAFs onto patho-gels.

We next questioned if the patterns of fiber alignment in CDMs correlate with the observed substrate stiffness-induced biphasic cell morphologies. For this, CAFs were cultured at high density on physio-gels, patho-gels or glass coverslips and allowed to produce CDMs. To assess the level of anisotropy at the cell-substrate interface, CDM fiber alignment was calculated from reconstructed quantitative confocal scanning images using the OrientationJ plugin for ImageJ software. For each condition, fiber alignment was calculated as the percentage of fibers oriented within 15 degrees from the measured orientation angle mode [33, 34]. Results revealed a biphasic fiber anisotropy: on physio-gels only ~30% of the fibers were aligned, while patho-gels induced a greater alignment (> 60%), yielding considerable fiber anisotropy. On glass coverslips, only ~40% of the fibers were aligned (measuring at the cell-substrate interphase). These data suggest a positive correlation between the underlying surface-instructed cell morphology and CDM fiber anisotropy (Figure 2A). A similar trend, albeit to a lesser degree, was observed using control fibroblasts (Supplemental Figure 1B).

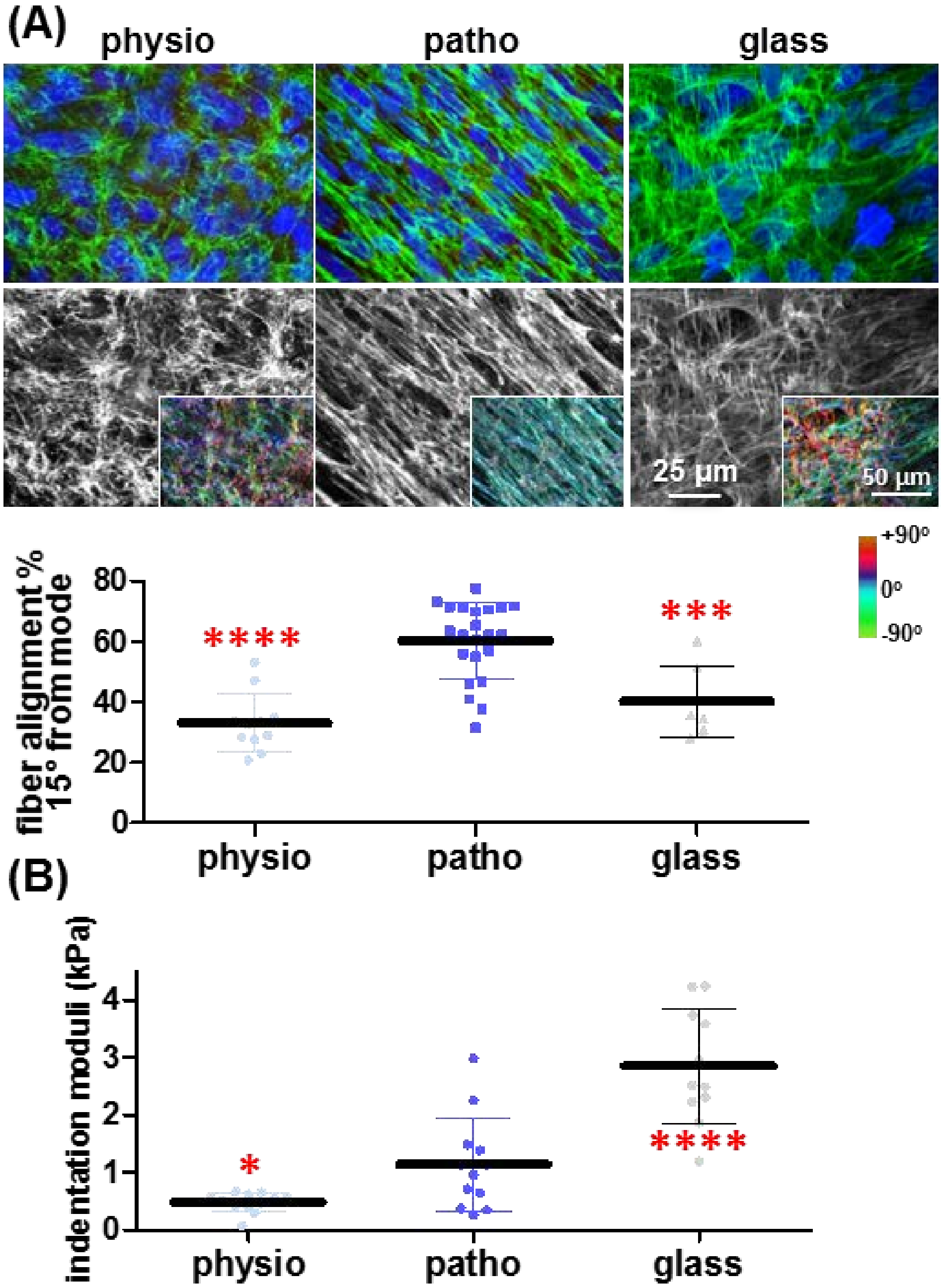

**Figure 2: Changes in substrate stiffness dictate a biphasic distribution of CDM alignment and a linear increase in CDM stiffness.** (**A**) Reconstituted maximum projections of confocal images obtained from gel or glass to CDM interphases, as

indicated. Fibronectin is shown in green merged with CAF cell nuclei (blue; top images) or as a monochromatic image (white; bottom images) containing a colored insert, indicative of fiber angle distributions, obtained with the 'OrientationJ' plugin of Image-J software and that was normalized using the hue function of Photoshop to show the cyan color as indication of fibers oriented on the mode measured angles. Color gradient bar on the right provides a color palate indicative of angle distributions. Graph includes the measured percentages of fibers distributed at 15˚ from the mode angle. Note that the CDM fiber anisotropy is at peak levels when CDMs are produced onto patho-gels. (**B**) Indentation moduli of decellularized CDMs produced onto the indicated substrates. Results are presented as mean ± standard deviation. Asterisks denote the following significance compared to measurements obtained using the patho-gel condition: * $p<0.05$, ** $p<0.01$, *** $p<0.005$ and **** $p<0.001$.

Interestingly, either on physio- or patho-gels, CAFs required only 3-5 days of culture to generate homogenous CDMs of >15 microns in thickness. This is less than half of the 8 days required to obtain CDMs on glass coverslips [33-35] with a CDM thickness of > 5 microns and an overall combined fiber anisotropy of > 55% (Supplemental Figure 2A). Similar results were obtained using additional patient-harvested CAFs (Supplemental Figure 2B). Taken together, our results suggest that both the intrinsic mechano-chemical coupling parameter of cells (i.e., CAFs and, to a lesser extent, in control fibroblasts) and the extrinsic effects imparted by the underlying substrate stiffness may modulate the level of CDM anisotropy.

Using atomic force microscopy (AFM) to gauge CDM vs. underlying gel stiffnesses, we first confirmed the indentation moduli of the engineered gels. As designed, gels exhibited the appropriate physiological and pathological stiffnesses of ~1.5 kPa and ~7 kPa, respectively. Interestingly, CDMs presented with numbers of magnitude lower stiffnesses (i.e., indentation moduli), compared to the corresponding underlying substrates (Figures 2B and Supplemental 3). Nonetheless, when control fibroblasts as opposed to CAFs were used to produce cell-derived ECMs, the indentation moduli were ~1 kPa in all cases, regardless of substrate stiffness (Supplemental Figure 3). These results suggest that, in terms of the indentation moduli, the extrinsic effects imparted by the various substrates mostly affect matrices produced by CAFs. Furthermore, the substrate-directed CDM stiffness is independent of both fiber anisotropy and cell-aspect ratios. Altogether these results suggest that, by using substrates of pancreatic physiological stiffness (i.e., physio-gels of ~1.5 kPa), it is possible to modulate the CDMs production, with regards to both fiber alignment and stiffness, to phenotypically resemble normal ECMs.

In order to explain the biphasic underlying stiffness-dependence of both CAF cell aspect ratios and corresponding CDM fiber alignments, we applied and modified an existing mechanochemical mathematical “contractile cell model” that was previously used to explain the correlation between substrate stiffness and cell polarization [32]. This model described how cells assume energy-favorable morphologies by "sensing” underlying surfaces of variating stiffness via intrinsic chemical energy, arising from myosin motors, and extrinsic mechanical energy imparted by substrate stiffness. Here we used a similar approach to include the cell-substrate interfacial energy that is directly associated with cell shape to determine the aspect ratios needed for cells to attain a given shape by minimizing their total free energy. In line with our experimental results, this modified model predicted a biphasic distribution of cell aspect ratios as a function of increase in substrate stiffnesses. Figure 3A depicts a theoretical cell that is cultured on a surface (blue), where the cell’s aspect ratio $(f)$ is defined as $f = a/c$ and in which $(a)$ is the cell’s length and $(c)$ is its breadth (i.e., with). The accompanying graph depicts the hypothetical biphasic changes in cell-aspect ratios predicting CAF shape changes with a maximum aspect ratio found at a stiffness that is above the physiological stiffness of the normal pancreas. These theoretical calculations predict and validate the experimental observations suggesting a) that, to minimize their total free energy, CAFs can dynamically alter their acquired aspect ratios in response to changes in the stiffness of their underlying substrates, and b) that these alterations involve both intrinsic/chemical and extrinsic/mechanical energies.

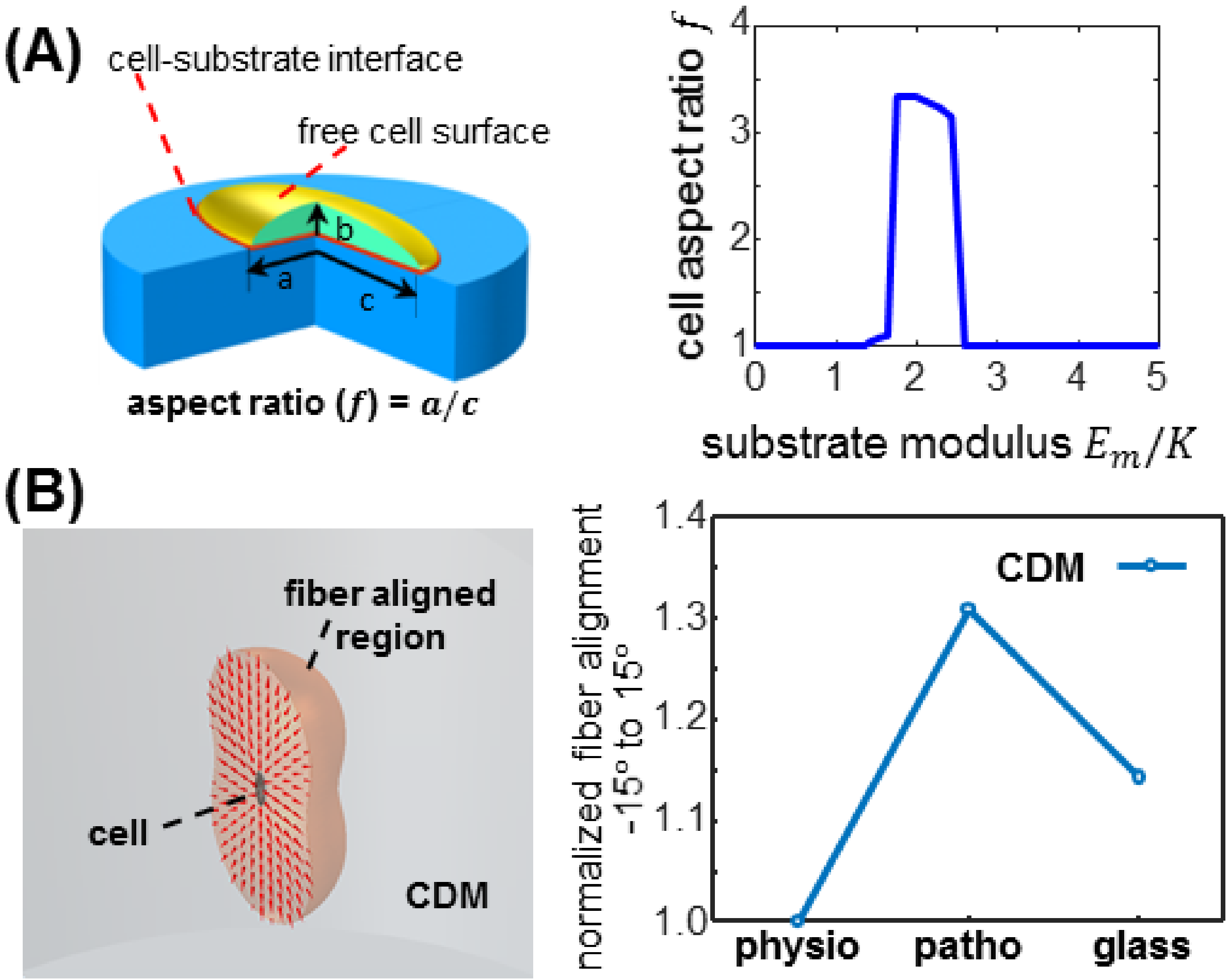

**Figure 3: Biphasic CDM alignment is explained in a minimum free energy model informed by substrate stiffness induced CAF aspect ratio.** (**A**) Depiction of the mathematical model conducted to explain the experimentally observed biphasic distribution in CAF aspect ratios. (**A**; left) Schematic representation of a cell (green) spread onto a 2D substrate (blue) used to calculate interface energy contribution to total energy. Both cell-substrate interface (red) and free cell surface (gold) are modeled as isotropic surfaces with surface energy γ_CM and γ_C (γ_C>γ_CM), respectively. (**A**; right) The minimum shape energies show biphasic response to changes in substrate modulus. K is the effective passive stiffness of the theoretical cellular actin network. (**B**) Depiction of mathematical model designed to explain the observed biphasic distribution in CDMs produced onto surfaces of increased stiffnesses, based on the observed/measured CAF aspect ratios. (**B**; left) Schematic depiction of the model suggesting that fiber alignment induced by single cell contraction: Light red area shows the spread of ECM fibers that are affected by a single cell, shown in the center. Red cones indicate the predicted local fiber orientations based on the center cell's aspect ratio. (**B**; right) Graph presenting the predicted CDM fiber alignment expected when informed via the measured CAF aspect ratios obtained in Figure 1. Please refer to Supplemental Table for additional data.

We next simulated the active crosstalk between CAFs and the underlying substrate to predict the effects of cell contraction on the initial fiber alignment. This simulation was informed by the measured cell aspect ratios and conducted by integrating a previously published model for fibrous ECM alignment [36] and the above depicted contractile cell models. Figure 3B shows that cell contraction prompts fiber alignment in a relative large region (gray), which is approximately 300 times that of the area of a cell. The red cones in Figure 3B indicate the predicted fiber orientation based on the cells' measured aspect ratio. Both model predictions and experimental results show that cells become elongated when cultured on surfaces of intermediate stiffness, indicating that contraction is mostly uniaxial when the cells are cultured on the patho-gels, with CAFs (as opposed to control fibroblasts) exhibiting the highest aspect ratios (Figures 1, 2 and Supplemental 1). These results suggest that the observed biphasic aspect ratios, informed by the extrinsic stiffness of underlying substrates combined with the intrinsic contractility (i.e., chemical energy) of CAFs, dictate the observed biphasic anisotropy of CDMs via maintenance of low free energy.

Our studies, as well as studies by other groups, have demonstrated that, while CDMs (i.e., produced by CAFs) are tumor-permissive, e.g., by supporting cancer cell growth and invasion, ECMs derived from normal fibroblasts are tumor-restrictive, e.g., regulate transcription, limit cell motility and metastatic invasion as well as alter the manner in which tumor cells transmit integrin-dependent biochemical signals [37-40]. Hence, we asked whether CDMs produced by CAFs growing on physio-gels, but not on patho-gels, could functionally restrict tumorigenic cell behaviors. For this, CDMs [34] produced onto physio- vs. patho-gels were decellularized and used as substrates on which we cultured transformed human pancreatic ductal epithelial cells (i.e., cells that were immortalized via human telomerase reverse transcriptase (hTERT) overexpression and transformed with oncogenic $KRAS^{G12D}$, concomitant with inactivation of tumor suppressors Rb and p53 [41]). CDM produced onto physio-gels limited cell proliferation, assessed by nuclear detection of Ki67 levels in these cells, by ~2 fold compared to levels attained on CDMs produced on patho-gels (Figure 4A). Used as controls, isogenic immortalized (e.g., hTERT only) benign human pancreatic epithelial cells presented with similar proliferative trends, albeit to a lesser extent (Supplemental Figure 4A). When both cell types were tested using CDMs made on coverslips and compared to matrices made onto gels, proliferation levels correlated with measured anisotropic fiber levels, yet the benign cells presented with reduced levels compared to transformed (KRAS-driven tumorigenic/invasive) cells (Supplemental Figure 4A). Bare gels (i.e., lacking CDMs) of pathological stiffness induced Ki67 incorporation/proliferation to levels akin to the ones observed on both cell types on glass, while physiological bare gels significantly restricted Ki67 incorporation in all cases (Supplemental Figure 4A).

Since anisotropic ECMs have also been shown to promote tumor cell invasion *in vitro* and *in vivo* [38-40, 42], we tested if CDMs generated on physio-gels could also restrict the invasive behavior of tumorigenic KRAS cells [41]. For this, we cultured pre-made KRAS cell spheroids, for 24 hours, onto CDMs produced on physio- vs. patho-gels and measured the area covered by invasive KRAS cells spreading. In line with our hypothesis, we observed that areas of invasive spreading decreased by ~2 fold when the spheroids were cultured on CDMs produced on physio-gels, compared to areas of cells spreading into CDMs that were produced onto patho-gels (Figure 4B). As controls, the same spheroids were cultured using all assorted matrices and 2D substrates. As seen in Supplemental Figure 4B, control fibroblastic-derived ECMs played a restrictive role in all cases; restricting area spreads similarly to invasive spreads attained by KRAS cells cultured in CDMs produced onto physio-gels. These data suggest the possibility that ECMs produced by control "normal" fibroblasts are inherently restrictive regardless of the substrate used to produce them. Importantly, similar results to the ones obtained with KRAS cells were also seen using an additional and well-known PDAC cell line, Panc1 (Supplemental Figure 4C). Taken together, our data suggest that biomechanical manipulations of CDMs, which restore a physiological stiffness-induced isotropic CDM organization/topology, can effectively restrain tumorigenic cell growth and invasion to levels like the ones observed when naturally restrictive ECMs were used.

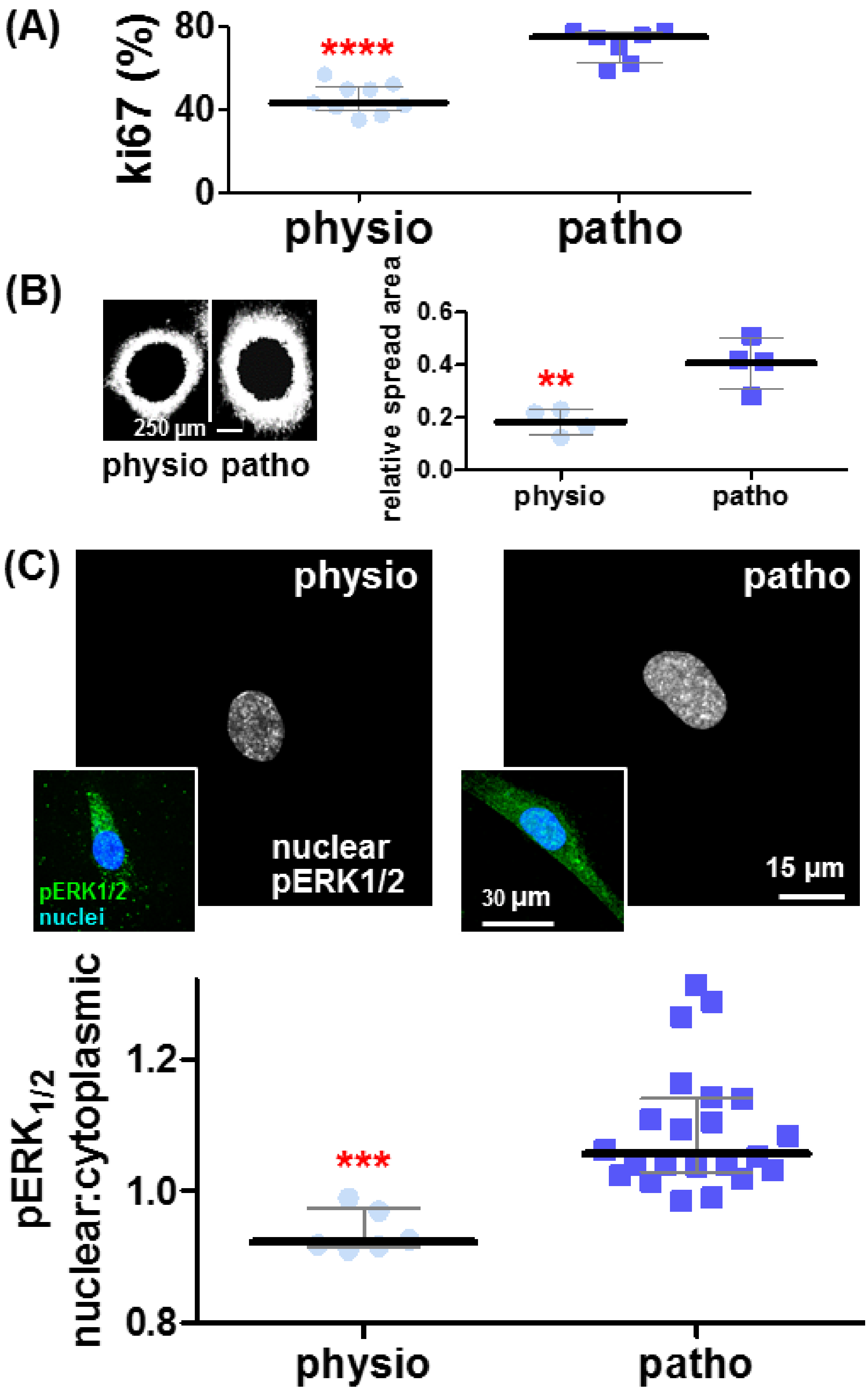

**Figure 4: CDMs generated on physiological substrates restrict tumorigenic cell behaviors by excluding pERK1/2 from the tumor cell nucleus.** KRAS-driven human pancreatic tumorigenic/invasive cells [41] were cultured in CDMs produced onto physio- vs. patho-gels. **(A)** Graph depicts measured percentages of nuclear Ki67 positive cells, showing median ± interquartile ranges. (**B**) Epifluorescence microscopy images, depleted of the original time 0 spheroid core, showing area spread of RFP expressing tumor cell spheroids at 24 hrs. (**B**-Graph) area spreads corresponding to 95 percent of RFP-spheroid measured intensities. The y-axis depicts relative area spread, which was calculated by dividing the final area spread by the area measured at time 0 hr. **(C)** Indirect immunofluorescence depicting monochromatic images of nuclear pixels (selected using the nuclear channel) of pERK1/2 staining, generated using SMIA-CUKIE 2.1.0. https://github.com/cukie/SMIA. Inserts include merged image of pERK1/2 (green) and nuclei (blue) channels. (**C**-Graph) SMIA-CUKIE 2.1.0. https://github.com/cukie/SMIA generated pERK1/2 nuclear localization intensity levels divided by cytosol localized pERK1/2 levels. Asterisks denote significances of: ** p<0.01, *** p<0.005 and **** p<0.001 compared to nuclei:cytosol ratios obtained when using CDMs produced on patho-gels. Please refer to Supplemental Table for additional data.

Nuclear accumulation of phosphorylated ERK1/2 (pERK1/2) is regarded as a surrogate downstream to constitutive KRAS signaling. Recent studies indicate that ERK2, rather than ERK1, is predominantly associated with the regulation of tumor cell invasion in 3D [43-48]. Hence, we next questioned the ability of CDMs produced onto physio- vs. patho-gels to maintain high pERK1/2 levels in general and, in particular, high nuclear pERK1/2. Western blotting revealed no difference in pERK1/2 levels in KRAS cells cultured on CDM produced on either physio- or patho-gels, yet there was a modest increase in pERK1/2 levels when KRAS-cells were cultured on CDM produced on glass **(**supplemental Figure 5A). Importantly, compared to CDMs produced on patho-gels, the nuclear localization of pERK1/2 was modestly, yet significantly (p=0.0003), reduced (from 1.1 to 0.9; by 18 %) in cells cultured on CDMs produced on physio-gels (Figure 4C). ECM controls, testing all experimental matrices and bare gels, showed a similar trend (supplemental Figure 5B-C and supplemental Table). This data suggests that in tumorigenic/invasive cells, nuclear localization of pERK1/2 is controlled via alterations in the CDM that, in turn, are affected by fine-tuning the underlying substrate stiffnesses. To elucidate the role of the two forms of ERK, we compared the effects of U0126, an inhibitor that indirectly blocks both ERK1 and ERK2, to the effects obtained by specifically knocking down ERK1 vs. ERK2 expression (Figure 5). As seen in Figure 5A, U0126 reduced CDM-induced invasion/spreading of tumorigenic KRAS cells by ~40%, compared to the invasion/spreading measured using vehicle control. To transiently knock down either ERK1 and/or ERK2, we used specific siRNAs and, as controls, equal amounts of a scrambled (e.g., non-specific) siRNA. As seen in Figure 5B, knocking down ERK2, but not ERK1, decreased anisotropic CDM-induced invasion by ~60%. Interestingly, invasion of human KRAS cells on CDMs produced on patho-gels under ERK1/2 or ERK2 blockage was similar to the invasion observed on both restrictive CDMs produced on physio-gels and by all the control fibroblast-derived ECMs (compare results in Figures 4, 5 and supplemental Figure 4). To assure that the CDM-induced effects were indeed ERK2 dependent, we verified that the level of phosphorylation levels of p90RSK, a downstream target of ERK2, was altered in accordance with pERK2, but not pERK1, changes (Figure 5). Taken together, our results suggest that ERK2, perhaps via phospho-p90RSK, is essential for matrix-induced invasion of human pancreatic tumor cells and that ERK2 blockage reduces anisotropic CDM-induced invasion.

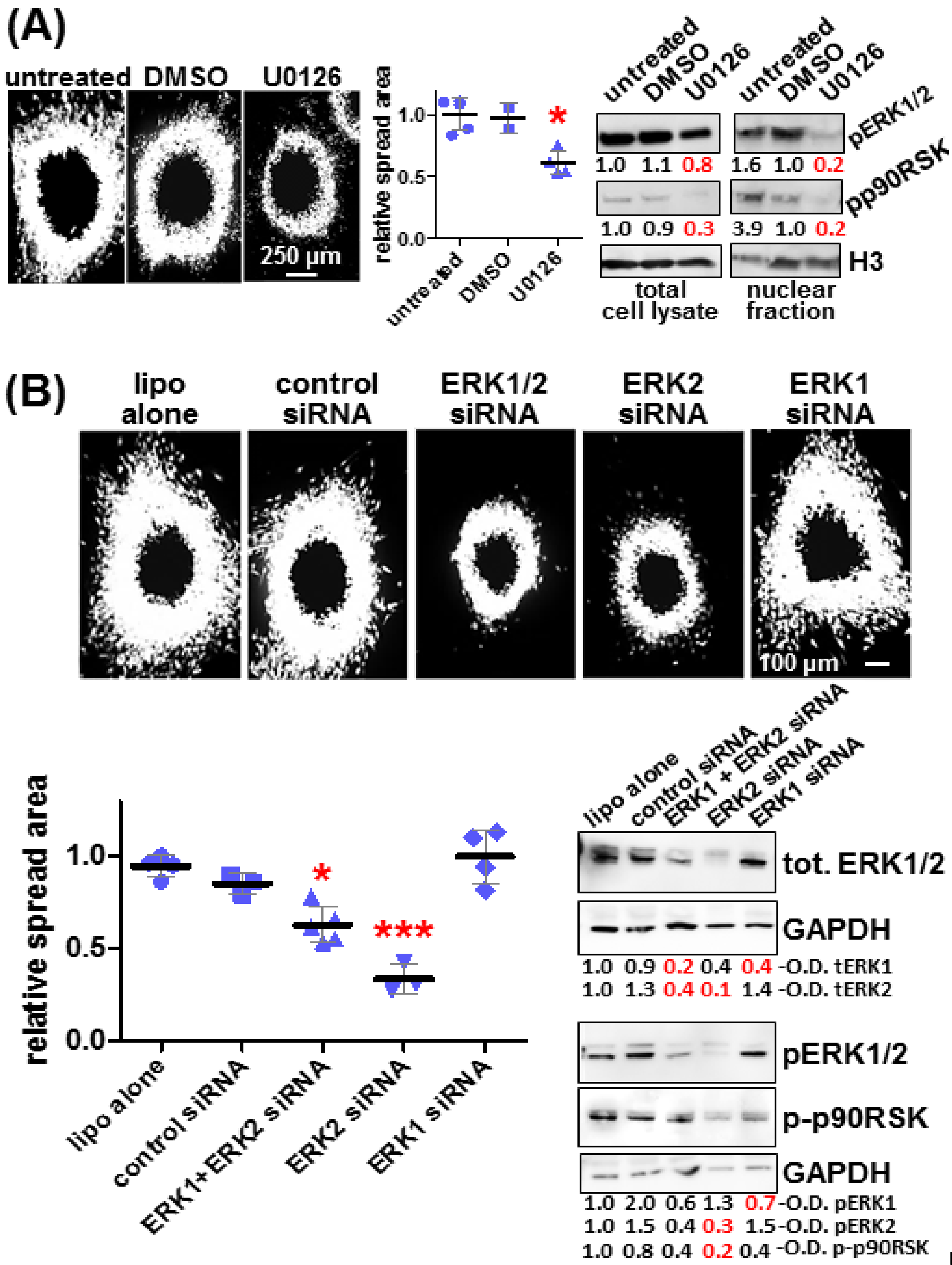

**Figure 5: CDM-induced PDAC invasion rates are regulated by tumoral pERK2**. (A) Tumor cell spheroids invading through CDMs produced onto patho-gels were treated with 20 µM of U0126 to inhibit MEK1/2 upstream to pERK1/2. Invasive area spreads were measured (graph) as before as well as in comparison to untreated and vehicle treated (DMSO) controls. Blots depict representative levels of pERK1/2 and phospho-p90RSK, downstream to pERK2, in total cell lysates (left) and in nuclear fractions (right). Representative measured optical densitometry (O.D.) are provided. (B) Same experiments as in A, but this time tumor cells were transfected with lipofectamine (lipo) or scrambled controls as well as siRNAs to ERK1 and/or ERK2 as labeled on the figure. Note how effects are seen when ERK2 is downregulated. Significance was calculated in comparison to untreated controls and asterisks denote * p<0.05 and *** p<0.005.

Lastly but importantly, to further validate both mathematical predictions and the *in vitro* findings, we assessed the nuclear localization pERK1/2 in 8 matching normal (e.g., physiological) and tumor (e.g., pathological) human pancreas samples. For this, we used our recently published simultaneous multiplex immunofluorescent approach and accompanying software needed for quantitative digital imaging analysis [33]. Results, shown in Figure 6A, demonstrated a ~3-fold increase in levels of nuclear localization of pERK1/2 in human PDAC samples, compared to nuclear pERK1/2 measured in matching normal/non-pathological pancreatic epithelial cells.

Taken together, our results suggest the possibility that the tumorigenic, desmoplastic stroma of human pancreatic cancer may be susceptible to modifications resulting in a tumor-restrictive stroma. This altered microenvironment might limit the nuclear localization of epithelial pERK1/2, which in turn could restrict KRAS-driven PDAC tumorigenicity (Figure 6B).

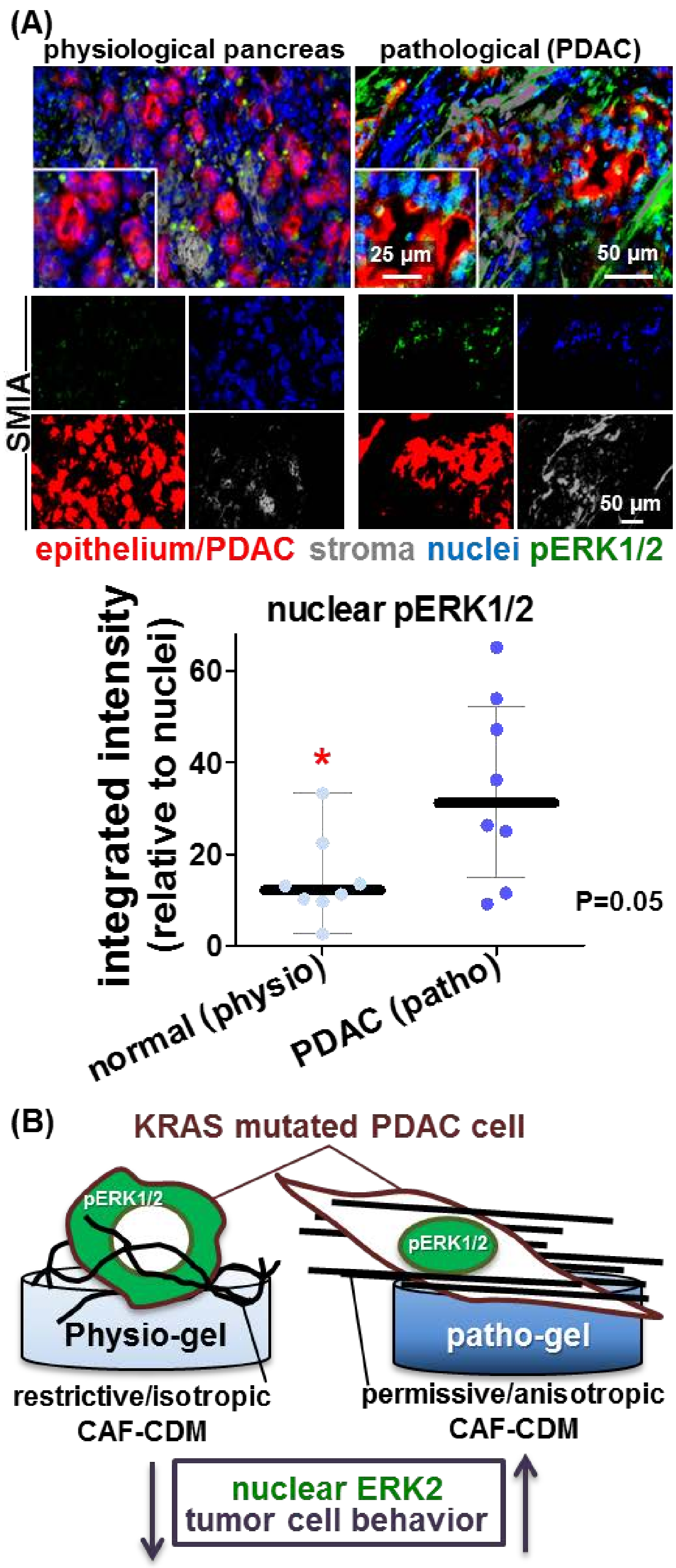

**Figure 6: Human normal stroma maintains pERK1/2 away from nuclei of pancreatic epithelial.** (**A**) Representative examples of pancreatic normal (physiological pancreas) and matched pathological PDAC formalin-fixed paraffin-embedded, FFPE, tissue samples stained and analyzed using the SMI approach and accompanying SMIA software [33]. Top panels correspond to merged images of a representative matching normal and PDAC sample showing epithelium/tumoral cells (red), nuclei (blue), stromal cells (grey) and pERK1/2 (green). The monochromatic panels shown below indicate "area masks" generated by the SMIA-CUKIE software (SMIA) and depict the levels of pERK1/2 (green) located solely at epithelial or tumoral (red) nuclei areas. Grey masks represent stromal areas. Graph summarizes the measured integrated intensity levels (obtained using SMIA-CUKIE) of nuclear tumoral (PDAC; patho) or normal epithelial (normal; physio) localized pERK1/2 levels. P value is indicated. (**B**) Summary cartoon to highlight that isotropic CDMs, obtained on a physiologically soft substrate, direct pERK1/2 (green) away from nuclear locations; restricting KRAS-driven tumorigenic/invasive, akin to PDAC, cell behaviors.

## Discussion

Pioneering work by the late Dr. Patricia Keely demonstrated that tumor-associated stromal ECM architecture is altered and that stromal ECM anisotropy *in vivo* serves as a prognostic indicator of poor cancer patient outcomes [49]. ECM-imparted physical cues dictate directional migration, as migrating cells often use the ECM fibers as attachment points during invasion [42, 50]. Hence, aligned (anisotropic) ECM fibers are a characteristic signature for PDAC-associated desmoplasia and indicative of poor patient prognosis [4, 5, 51]. Ablation of PDAC-associated desmoplasia was tumor-promoting pre-clinically and detrimental to patients in clinical trials [21, 52, 53]. Hence, approaches that can modulate the pro-tumorigenic aspects of desmoplasia (e.g., desmoplastic ECM anisotropy) to harness the naturally tumor-restrictive features of a normal microenvironment are highly sought by the field. As such, genetic and biomechanical regulation of CAFs seeks to alter their ability to contract and produce an anisotropic ECM. For example, increased levels of stromal caveolin-1 in fibroblasts were shown to impact stromal ECM anisotropy, via p190RhoGAP activation, which increases intrinsic cell contractile forces and, in turn, facilitates *in vivo* metastatic escape [39]. The presence of syndecan-1 in stroma alters the ECM architecture, *in vivo* and *in vitro*, and induces the production of an anisotropic ECM that promotes cell invasion [54]. Further, overexpression of fibroblast activating protein (FAP) in naive fibroblasts prompts the formation of anisotropic ECM, akin to pancreatic CDM, and also induces increased PDAC cell invasion [40]. Alternatively, physical approaches have been used to modulate ECM anisotropy, however most systems rely on collagen gels or synthetic scaffolds which lack *in vivo* biochemical complexity. Thus, despite reports suggesting that ECM anisotropy is a key regulator of PDAC tumorigenesis [4, 5, 51], the particular biomechanical cues that enable production of tumor-restrictive ECM remain mostly elusive.

Substrate stiffness strongly influences a wide range of cell functions, including some with ECM-modifying properties, such as cell spreading, survival, proliferation, differentiation, and migration [55]. Substrate stiffness also enables ECM remodeling by modulating the expression of genes for ECM proteins and ECM-altering enzymes [56]. Fibronectin fibrillogenesis, which is necessary for collagen fibrillogenesis *in vivo* [57], is greatly affected by the underlying substrate stiffness [58]. Fibrillogenesis per se is necessary for altering cell-intrinsic forces [59]. Substrate stiffness can also drive collective cell migration, which is critical for some types of cancer invasion [60]. Nonetheless, not many studies have highlighted the role of substrate stiffness on CAF-derived matrix production, which is one of the most important factors believed to influence cancer progression [61]. To correctly interpret the role of biomechanical matrix properties that affect cell behavior, it is important to select a system that accurately mimics the relevant *in vivo* physiological and pathophysiological microenvironments. In the present study, we adapted our previously described, cell-derived ECM system [34, 62] to identify means to alter desmoplastic ECM anisotropy and the role that anisotropic ECMs could play in dictating pancreatic cancer tumorigenic cell behavior, such as proliferation and/or invasion. The advantage of this well-characterized system is that it is comprised of natural, cell-derived materials and that it offers an *in vivo*-like biochemical complexity [62]. However, one known technical limitation of using CDMs produced on glass coverslips is a gradient-like heterogeneity in fiber anisotropy, apparent between the bottom (at the glass ECM interphase) and top ECM fiber layers (Movie 1). Remarkably, using the physio and patho-gels as shock absorbing/buffering materials not only permitted fine-tuning of the underlying substrate stiffness and direct the anisotropy of CDM fibers (Figure 2), but it allowed doing so in a homogenous way (Supplemental Figure 2).

We developed a mathematical model using the experimentally measured cell aspect ratios (Figure 1 and Supplemental 1A) to predict the anisotropy levels of matrix fibers being produced at the cell-substrate interphase. Substrate stiffness affects cytoskeletal actin organization, which, in turn, regulates numerous cellular functions, such as cell spreading and migration [63]. Because of the organization of the actin network, cells contract along their long axes and generate aligned ECMs [36]. This effect is increasingly pronounced as the cell becomes more elliptical [30, 32]. Interestingly, intermediate stiffness (e.g. 7 kPa) is associated with induction of maximum elongation of cell and migration, whereas softer stiffness (e.g. 1.5 kPa) is associated with dramatic reduction in cell spreading [64, 65]. This prompted us to hypothesize that CAFs cultured on soft substrates may exhibit “normalization” of ECM anisotropy. Our published model, also referred to as “constitutive material model” [36], is a continuum model that addresses fiber alignment and long-range force transmission (i.e., the ability of cells to “sense” each other at long distances within an ECM-like mesh within matrices generated on physiological versus pathological environments). The model also captures the nonlinear elastic behavior of matrices in response to cellular contraction. Based on our experimental data (Figures 1 and 2) and supported by mathematical modeling (Figure 3), we suggest that extrinsic (e.g., substrate stiffness) and intrinsic forces (e.g., ECM producing cell’s contraction ability) collectively affect the aspect ratios of CAFs, which in turn control CAF-derived ECM anisotropy. Therefore, we propose changing substrate stiffness as novel means for manipulating extrinsic cues, which may influence the ability of human PDAC-associated CAFs to biomechanically remodel CDMs by altering their key biomechanical properties: anisotropy and indentation moduli. Our data (Figures 3 and 4) further demonstrated that tumor restrictive responses, particularly loss of invasion, are highly correlated with matrix anisotropy. This study is in line with the previous data that ECM anisotropy is a major predictor of tumor responses, particularly invasion [30, 66, 67].

Despite harboring “permanent” KRAS mutations, PDAC cells require stromal cues to effectively trigger and maintain constitutive KRAS activity *in vivo* [13]. High, constitutive epithelial/tumoral levels of ERK1/2 activity constitute a surrogate to assess high KRAS activity in PDAC cells [68]. ERK1/2-regulated epithelial to mesenchymal transition has been linked to poor PDAC survival [69]. Activation of key mechano-transducing molecules such as YAP, which depend on ERK1/2 phosphorylation, is also a trademark of PDAC [70]. In this study, we observed sustained high levels of nuclear pERK1/2, as opposed to merely increased cytosolic levels, in KRAS-driven tumorigenic/invasive human PDAC cells cultured on anisotropic matrices (i.e., CDMs produced onto patho-gels). Yet, pERK1/2 was effectively excluded from the nucleus when the cells were cultured on isotropic matrices (i.e., CDMs produced onto physio-gels). Additionally, ERK2, and not ERK1, was responsible for regulating the observed ECM-induced PDAC cell responses (e.g., invasion in Figure 5). These data are in line with our own early studies and studies by others which have highlighted a role for ERK2 in tumorigenic responses to extracellular-imparted cues [43-48]. Interestingly, high phosphorylated ERK2 levels have been correlated with poor survival [71]. Specific nuclear localization of ERK2 induces epithelial mesenchymal transition [70] and contributes to drug resistance [72]. Our data suggest that pERK2, possibly regulated by its canonical effector p90RSK, is implicated in ECM-induced oncogenic KRAS-driven PDAC invasion (Figure 5). This observation agrees with reports showing synthetic lethality between ERK1/2 inhibition and p90RSK or its downstream effector CDC25C [73]. It has been reported that integrin activity is needed for ERK1/2 nuclear translocation [74]. Future work will focus on assessing if, by remodeling the CDM architecture, we also alter integrin signaling.

Lastly, this study demonstrated that human PDAC tissues present with enriched tumoral nuclear pERK1/2 while normal human tissues exhibit lower levels of nuclear pERK1/2 in the pancreatic epithelium (Figure 6). While increased overall levels of pERK1/2 have been reported in human PDAC [75], only few studies have looked at epithelial/tumoral nuclear pERK1/2 localization [76]. Therefore, these findings may have important implications regarding the therapeutic use of drugs that could exclude pERK1/2 from tumoral nuclei. Future efforts will be directed towards further evaluating mechanisms on how ERK2, enabled via anisotropic CDMs, may control pERK1/2 translocalization to PDAC nuclei.

**Conclusions:**

In this study, we have demonstrated that CAFs can be enticed to generate pancreatic cancer-restrictive CDMs, provided the underlying substrate rigidity matches that of a physiological pancreas. We propose that the *in vitro* measured results can be modeled mathematically, informed by the substrate stiffness' extrinsic forces combined with CAFs' intrinsic contractility, which jointly directs a biphasic matrix fiber anisotropy by maintenance of a low free energy. We found that CDMs generated by CAFs onto physiologically soft gels are tumor-restrictive and limit Ki67 incorporation, indicative of reduced rates of proliferation and invasion of oncogenic KRAS driven pancreatic cancer cells. Our results also suggest that loss of nuclear pERK1/2 in cells cultured on isotropic CDMs is probably regulated by restricting ERK2 activity. The observed *in vitro* results correlated with *in vivo* measured epithelial nuclear pERK1/2 levels, supporting the validity of our tunable pathophysiological 3D CDM system. Hence, the therapeutic reprogramming of stromal ECM and/or targeting tumoral ERK2 may provide future means to contain PDAC, and possibly other KRAS-driven neoplasias.

**Material and Methods**

**Contact for Reagent and Resource Sharing**

All information (manufacturer/source) regarding the chemical and biological reagents has been compiled in the Reagents Table. Further information and requests for resources and reagents should be directed to and will be fulfilled by the Lead Contacts, Drs. Edna Cukierman (Edna.Cukierman@fccc.edu) and Peter I. Lelkes (pilelkes@temple.edu).

**Experimental Model and Subject Details**

**Cell Lines**

Human pancreatic CAFs were isolated using an Institutional Review Board approved protocol. Cells were characterized, immortalized and authenticated as previously described [77, 78]. Since cultured NIH-3T3 fibroblasts (from ATCC, CRL-1658™) do not undergo spontaneous myofibroblastic activation [35], these cells were used as inactive fibroblastic cell controls as before [38]. All cells were maintained in a humidified incubator at 37°C and 5% $CO_2$. Control fibroblasts and CAFs were cultured in Dulbecco's Modified Eagle's Medium (DMEM; from Mediatech (Manassas, VA) supplemented with 10% FBS, 100 U/mL Penicillin, 100 mg/mL Streptomycin and 2 mM L-Glutamine. Isogenic pancreatic ductal epithelial cells, hTERT and KRAS (hTERT/E6/E7/KRAS; CRL-4038™ [41]) were from ATCC and cultured in growth medium containing four parts of low glucose DMEM and one part M3 supplemented with 5% FBS containing 100 U/mL Penicillin and 100 mg/mL Streptomycin.

**Method Details**

**Preparation of Polyacrylamide gels**

Circular glass coverslips, 18 mm in diameter (Carolina Biological Supply Company; Burlington, NC), were activated using 3-Aminopropyl triethoxysilane (APTES) for 10 minutes and washed extensively with distilled water followed by treatment with 0.5% glutaraldehyde for 1 hr. To prepare the gel solutions, acrylamide and N,N′-Methylenebisacrylamide solution were mixed together in distilled water in the desired ratios to generate gel precursor solutions for predicted Young's moduli of ~1.5 (physio-gels) and ~7.5 kPa (patho-gels) [79]. The final percentage of gel solutions for a ~1.5 kPa gels was 3% acrylamide and 0.15% bisacrylamide and 10% acrylamide and 0.1% bisacrylamide for a ~ 7.5 kPa gel. Gel stiffness were validated using atomic force microscopy. Gel polymerization was initiated by addition of crosslinkers 10% w/v APS and N,N,N′,N′-Tetramethylethylenediamine accelerator (TEMED) at dilution 1:000 and 1:10,000 respectively from their stock solutions. After gentle mixing, 120 µl of the gel solution were pipetted onto the activated coverslips and a dichlorodimethylsilane -treated coverslip was carefully placed on top of the gel solution. Gels were allowed to polymerize at room temperature for ~10-15 min. The top glutaraldehyde and dichlorodimethylsilane (DCDMS)-treated coverslip was gently lifted and gels were washed with Milli-Q water and sterilized under a UV lamp ( 365 nm) for 15 min. Covalent conjugation of gels with 50 µg/ml collagen-I was performed in 50 mM HEPES buffer, 8.5 pH. Collagen-I was crosslinked to the gels using Sulfo-SANPAH for 15 min under the UV lamp, as above. Collagen-coated gels were washed extensively with PBS and stored in PBS at 4°C for up to two weeks. The Collagen-conjugated gels were equilibrated for 30 min with culture media at 37°C prior to seeding with the various fibroblasts.

**Preparation of CDMs onto polyacrylamide gels**

Glass cover slips containing gels coated with collagen were placed inside wells of a 24 well cell culture plate (gel side up). Pyrex® cloning cylinders (Sigma Aldrich, St. Louis, MO), 8 mm (height) x 8 mm (diameter), were carefully placed at the center of gels and 100 µl culture medium containing $4x10^4$ CAFs (or control fibroblasts) was carefully pipetted inside the cylinders. Cylinders were removed after ~1 hr. to allow the cells to attach. Next, the cells were covered with 1 ml of fibroblast culture medium, listed above, supplemented with 50 µg/ml ascorbic acid. A similar procedure, omitting the use of cloning cylinders, was employed for producing CDMs by seeding the cells (CAFs and normal controls) directly on collagen coated glass coverslips, as previously published [77]. Fort his, $1.25x10^5$ cells per 12 mm coverslips were plated. In both cases, media including freshly weighted and diluted ascorbic acid (50 µg/ml) was added every day except on the last day (e.g., before extraction).

Following matrix production (3 days for CDMs on gels or 8 days for CDMs on glass), decellularized matrices were obtained using an alkaline detergent (0.5% Triton X-100 and 20 mM $NH_4OH$ in PBS), followed by DNase I (50 U per mL) treatment [77]. The resulting decellularized matrices were washed three times with PBS and stored at 4°C for up to 2 months. All decellularized matrix batches, of control fibroblast ECMs and CAF CDMs, underwent rigorous quality control as published [77].

**Indirect immunofluorescence and image analysis**

Indirect immunofluorescence was as previously described [77]. Briefly, samples were fixed/permeabilized, for 3 minutes, with 4% (w/v) paraformaldehyde (EM-grade from Electron Microscopy Sciences), 0.5 % (v/v) Triton X-100, and 50 mg/ml sucrose in Dulbecco’s phosphate-buffered saline and continued fixing, in the absence of triton, for 20 minutes. Samples were blocked for 2 hr. with Odyssey blocking buffer (TBS) (LI-COR Biotechnology, NE, Cat. 927-50100). For matrix assessments, samples were incubated, for 1 hr., with a rabbit anti-mouse fibronectin antibody (25 μg/ml, Abcam, UK Catalog no: ab2413). After incubation with primary antibody, the matrices were washed three times, for 10 min. each, with tris buffer saline containing 0.5 % v/v Tween 20 (TBS-T buffer) and incubated at room temperature for 45 min with donkey anti-rabbit Cy5 conjugated secondary antibody (15 μg/ml, Jackson ImmunoResearch, PA Catalog no: 711-175-152). Samples were washed with TBS-T, three times. Nuclei were stained with SYBR green (1:50,000 dilution, Thermo Fisher Scientific, Waltham, MA) and samples were mounted as previously detailed [77]. Images were captured using a spinning disk confocal microscope (Ultraview, Perkin-Elmer Life Sciences, Boston, MA) equipped with a 60X (1.45 PlanApo TIRF) oil immersion objective. For each condition, three independent experiments were conducted and a minimum of 7 images per sample were obtained.

CDM (or control ECM) alignment measurements were conducted using ImageJ OrientationJ plug analyses as published [77, 78]. Isotropic CDMs were identified as matrices containing < 55% alignment, while anisotropy was identified as > 55% of fibers distributed at 15 degrees from the mode angle [77, 78].

For ERK1/2 subcellular localization quantification, KRAS transformed/invasive [41] cells were incubated with rabbit anti-human phospho- p44/42 ERK1/2 (Thr202/Tyr204) (Cell Signaling Technology, Danvers, MA, Cat no. 4370,) followed by Cy5-coupled secondary antibody and SYBR green, as above. Subcellular localization of p-ERK1/2 (cytosolic vs. nuclear) was quantified using our publically available software, SMIA-CUKIE 2.1.0 https://github.com/cukie/SMIA [78]. Images corresponding to the same experimental conditions and the same staining procedures were processed identically; 16 to 8 bit level conversions were conducted using identical parameters. To find suitable thresholds, to inform the SMIA-CUKIE software, and distinguish between signal and noise, we applied intensity histogram distributions obtained from Photoshop. Selected threshold values (between 1 and 254) for each staining (pERK1/2 or nuclei) were consistently used through the study. Images were sorted into experimental folder batches, using the “make a batch” software in https://github.com/cukie/SMIA, to include monochromatic matching images of nuclei and pERK1/2 per simultaneously acquired images, which served as inputs for the SMIA-CUKIE 2.1.0 software. Mean intensity levels of pERK1/2 in nuclei vs. cytoplasmic fractions were measured and outputs were plotted as nuclei:cytoplasmic ratios. Show in the figures are representative monochromatic image outputs, displaying positive pixels indicative of "nuclear" only pERK1/2 levels, accompanied by total pERK1/2 nuclei color overlays.

**Atomic force microscopy (AFM)**

The stiffnesses of the various (gel and cell-derived matrices) substrates were assessed by AFM-nanoindentation, carried out on a Dimension Icon AFM (BrukerNano, Santa Barbara, CA), using a custom-made microspherical tip. The colloidal probe used was generated by attaching a 5 µm-radius polystyrene microsphere (PolySciences, Warrington, PA) to the end of a tipples cantilever (Arrow-TL1Au, NanoAndMore USA, Watsonville, CA) using M-bond 610 epoxy (Structure Probe Inc., West Chester, PA). All tests were conducted using filtered 1× PBS to simulate a physiological fluid environment. The probe tip was programmed to indent into the sample at a constant z-piezo displacement rate of 5 µm/s, up to a maximum indentation depth ~ 1 µm. All CDMs (and control ECMs) used for this study were at least 8 µm thick, i.e. exceeding the minimum thickness of 7 µm required for CDM quality control [77]. Each sample was tested at a minimum of 10 randomly selected locations to ensure consistency and/or to account for spatial heterogeneity. The indentation modulus $E_{ind}$ was calculated by fitting the loading portion of each indentation force-depth curve to the Hertz model,

$$F = \frac{4}{3}\frac{E_{ind}}{\left(1-\nu^2\right)}R_{tip}^{1/2}D^{3/2} \quad (1)$$

where F is the indentation force, D is the indentation depth, ν is the poison's ratio (0.49 for highly swollen hydrogels) (57), and $R_{tip}$ is the radius of the probe tip (≈ 5µm). Since the thickness of the gels (>200µm) is more than 2 orders of magnitude greater than the maximum indentation depth, the substrate constraint effect was minimal, and thus, finite thickness correction was not needed. The comparison between CDMs and adjacent bare gels was done by probing regions with or without the ECM on the same gel for consistency. This was possible because CDMs were constrained to the cloning cylinder areas, with the adjacent bare gel areas serving as internal controls. The mechanical properties of the latter were indistinguishable from those of intact bare gels that had never been coated with CDMs.

**Mathematical model for predicting cell shape**

Cells change their shapes in accordance with the physicochemical properties of the underlying substrate [80]. To mathematically understand how substrate stiffness influences cell morphology, we consider a cell cultured on a 2D substrate. We use an energy criterion to determine the cell shape, i.e. we hypothesized that a cell adjusts its shape in order to minimize the total free energy of the cell-substrate system. The total free energy can be written as,

$$E = E_{cell} + E_{subst} + E_{int} \qquad (2)$$

where $E_{cell}$ is the cell energy, $E_{subst}$ is the elastic energy of the underlying substrate and $E_{int}$ is the interface energy (including the basolateral cell-substrate interface and the apical free cell surface). The cell energy is a function of the elastic energy (accounting for cell deformation) and the motor density (accounting for contractility). Based on the model for contractile cells [32], $E_{cell}$ can be written as,

$$E_{cell} = \int_{Cell} U_C(\varepsilon_{ij}^C, \rho_{ij})\, dV \qquad (3)$$

where $U_C$ is the cell energy density, $\varepsilon_{ij}^C$ is the elastic deformation of the cell and $\rho_{ij}$ is the motor density. The interface energy consists of the basolateral cell-substrate interface energy and the apical free cell surface energy.

$$E_{int} = \gamma_{CS} S_{CS} + \gamma_C S_C \qquad (4)$$

where $\gamma_{CS}$ and $\gamma_C$ are interface/surface energy density for cell-substrate interface and free cell surface respectively, $S_{CS}$ and $S_C$ are the area for cell-substrate interface and free cell surface respectively.

We characterized the cell shape by defining the aspect ratio $f = a/c$, where $a$ stands for length and $c$ is the cell breadth. For a given substrate (a fixed stiffness), we computed the total free energy of the cell-substrate system for various aspect ratios, and chose the energy-minimized one as the preferred cell shape. Next, we varied the substrate stiffness and obtained the hypothetical cell aspect ratio as a function of stiffness.

**Mathematical model for predicting CDM alignment**

In terms of the stress-dependent regulation of cell contractility, the contractile stress of the actin network can be written [32] as,

$$\sigma = \rho + K\varepsilon \qquad (5)$$

where $\rho$ is the density of force-dipoles (representing myosin motors/contractility) in the actin network, $\varepsilon$ is the strain of the actin network and $K$ is the effective passive stiffness of the actin network. The contractility itself depends on the mechano-chemical coupling through mechano-signaling pathways, such as Rho-Rock and myosin light chain kinase [32];

$$\rho = \frac{\beta\rho_0}{\beta-\alpha} + \frac{\alpha K-1}{\beta-\alpha}\varepsilon \qquad (6)$$

where $\rho_0$ is the contractility in the absence of adhesions, $\alpha$ and $\beta$ denote mechano-chemical coupling parameters. Additional details of this model have been described elsewhere [32].

**Short Interfering RNA (siRNA) Transfections**

Transient transfections were performed on the tumorigenic/invasive KRAS cells [41], using Lipofectamine® 2000, following the manufacturer’s instructions (Thermo Fisher Scientific, Waltham, MA). Non-targeting SMARTpool and siRNA targeting ERK1 or ERK2, each comprising four distinct siRNA species, were from Life Technologies-Dharmacon (Lafayette, CO). Transfections were carried out in KRAS cell growth media, without FBS or antibiotics. Cells were trypsinized, plated at a density of $1x10^5$ per well, in a 6-well plate, and mixed with transfection medium, as per manufacturer’s instructions. The plate was placed in the incubator for 5 hrs. Following incubation, media was replaced with regular KRAS cell growth media (as above) and cells were cultured for an additional 48 hrs. For spheroid spread experiments, cells were trypsinized 24 hrs. post transfection and used for spheroid formation, followed by spheroid spread assay (see below).

**Western blotting**

Cell lysates were obtained using a cell lysis buffer from Cell Signaling Technology (Catalog no. 9803, Danvers, MA) supplemented with Pierce$^{TM}$ Phosphatase and Protease Inhibitor Mini Tablets (Cat no. 88667 and 88665, respectively) from Thermo Fisher Scientific (Waltham, MA). Proteins were resolved by 4-20% SDS-PAGE gels (Bio-Rad, Hercules, CA) at 60 V and transferred to PVDF membranes using semi-dry transfer (Bio-Rad, Hercules, CA). Protein transfer was carried out at 20 V for 30 minutes. Blots were incubated with the following primary antibodies: Rabbit anti-human Phospho- p44/42 ERK1/2 (Thr202/Tyr204) (Cat no. 4370) and rabbit anti-human total- p44/42 ERK1/2 (Cat no. 9102) from Cell Signaling Technology (Danvers, MA). Anti-phospho-p90RSK1 (Ser380) (Cat no. 04-418) and anti-human glyceraldehyde 3-phosphate dehydrogenase (GAPDH) (Cat no. MAB374) from Millipore (Billerica, MA). Horseradish peroxidase- conjugated, anti-species matched, secondary antibodies were from Sigma Aldrich (St. Louis, MO). Protein bands were visualized using the Protein Simple FluorChemE System, (San Jose, CA). For biochemical nuclear level assessments of pERK1/2 we used the Subcellular Protein Fractionation Kit (Thermo Scientific, Waltham, MA), according to the manufacturer's instructions.

**Cell proliferation assay (Ki67)**

Pancreatic human epithelial cells (benign hTERT immortalized and KRAS-driven tumorigenic/invasive; [41]) were plated at a density of $2x10^4$ cells/ml, per sample, and cultured for or 24 hrs. Cells were fixed as stated in **Indirect immunofluorescence and digital imaging analyses** section above, prior to staining with anti-Ki67 antibody (Cat no. ab15580, Abcam, Cambridge, UK), using the same protocols as above. The fraction of proliferating cells was determined by counting the number of cells stained positive for Ki67 divided by total number of nuclei, stained using Hoechst 33342 solution (Calbiochem, Billerica, MA). At least 5 images were taken per condition, a minimum of two samples was used for each experiment and experiments were performed independently a minimum of three times. Data from all three experiments was pooled and plotted.

**Lentiviral infection of KRAS cells**

Target cells were seeded at ~40% confluence in a 6 well plate and allowed to attach overnight. The next day, target cells were infected in the presence of media containing mCherry lentivirus (Plv-CMV-Puro vector) and 10 ug/ml polybrene (Santa Cruz). After 24 hrs. media was replaced with complete KRAS media and allowed to grow for an additional 48 hours. 72 hrs. after the initial infection, cells were screened for the presence of mCherry, using the EVOS microscope system. After confirmation of red color, media was replaced with KRAS media containing puromycin (12 ug/mL) and selection of mCherry positive cells occurred over the next 7-10 days. The resulting cells that survived selection were then used for subsequent experiments.

**Spheroid spreading invasion assay**

Red fluorescence protein (RFP)-expressing tumorigenic/invasive KRAS cells were trypsinized and resuspended in spheroid formation media (Irvine Scientific, Santa Ana, CA, Catalog ID: 91130) overnight; 30 µl drops containing $2.5x10^3$ cells were carefully placed on a lid of a sterile 100 mm Petri dish. The dish was filled with 5 mL media, to avoid condensation or drying, and the lid with the "hanging drops" was carefully placed, drops facing down, and incubated overnight. Spheroids were carefully removed from the lids and placed, one by one, onto the assorted matrices, gels and glass substrates and allowed to adhere, for 2 hrs. Subsequently, the spheroid formation media replaced by regular pancreatic cancer growth media. Invasion assays lasted 48 hrs. Areas of cell spreading were visualized in an inverted microscope equipped with epifluorescent image acquisition capabilities, using a 10X objective. Data were normalized to the initial size of each spheroid, as measured at time 0. When indicated, spheroids were treated overnight with 20 µM of the MEK1/2 inhibitor U0126 from Calbiochem or the equivalent volume of DMSO, or pre-transfected 24 hrs. prior to spheroid formation with the assorted siRNAs. Images were processed using MetaMorph 7.8.1.0 software (Molecular Devices, Downingtown, PA). A minimum of 5 spheres per condition were analyzed in at least three independent experiments.

**Quantification and Statistical Analysis**

All experiments included a minimum of duplicate samples and repeated independently at least three times. Data was plotted using GraphPad Prism and analyzed using unpaired Student’s t-test analyzing unpaired conditions each time. Values are presented as median ± interquartile range or mean ± standard deviation, as indicated in the figure legends. Asterisks depicting statistical significance are indicated, when relevant.

**Reagent Table**

| REAGENT or RESOURCE | SOURCE | IDENTIFIER |
|---|---|---|
| Antibodies | | |
| Rabbit anti-mouse fibronectin antibody | Abcam | Cat # ab2413, RRID:AB_2262874 |
| Donkey anti-rabbit Cy5 conjugated antibody | Jackson ImmunoResearch | Cat # 711-175-152, RRID:AB_2340607 |
| Rabbit anti-human phospho- p44/42 ERK1/2 (Thr202/Tyr204) | Cell Signaling Technology | Cat # 4370, RRID:AB_2315112 |
| Rabbit anti-human total- p44/42 ERK1/2 | Cell Signaling Technology | Cat # 9102, RRID:AB_330744 |
| Anti-phospho-p90RSK1 (Ser380) antibody | Millipore | Cat # 04-418, RRID:AB_673094 |
| Anti-Ki67 antibody | Abcam | Cat # ab15580, RRID:AB_443209 |
| Anti-GAPDH antibody | Millipore | Cat # MAB374, RRID:AB_2107445 |
| Biological Samples | | |
| Rat Tail Collagen-I | Thermo Fisher Scientific | Cat # A1048301 |
| Surgical samples of patient harvest pancreatic tissue (normal and cancer tissue) | Fox Chase Cancer Center Biological sample Repository | NA |
| Chemicals, Peptides, and Recombinant Proteins | | |
| U0126 | Calbiochem | Cat # 662005 |
| Sulfo-SANPAH | Thermo Fisher Scientific | Cat # 22589 |
| APTES | Sigma-Aldrich | Cat # A3648 |
| Glutaraldehyde | Sigma-Aldrich | Cat # G6257 |
| Acrylamide | Sigma-Aldrich | Cat # A4058 |
| N,N′-Methylenebisacrylamide solution | Sigma-Aldrich | Cat # M1533 |
| TEMED | Sigma-Aldrich | CAS Number 110-18-9 |
| APS | Sigma-Aldrich | Cat # A3678 |
| DCDMS | Sigma-Aldrich | Cat no. 440272 |
| Paraformaldehyde | Electron Microscopy Sciences | CAS #30525-89-4 |
| Odyssey blocking buffer (TBS) | LI-COR Biotechnology | Cat # 927-50100 |
| SYBR green | Thermo Fisher Scientific | Cat # S7563 |
| Lipofectamine® 2000 | Thermo Fisher Scientific | Cat # 11668027 |
| Cell lysis buffer | Cell Signaling Technology | Cat # 9803 |
| PierceTM Phosphatase | Thermo Fisher Scientific | Cat # 88667 |
| Protease Inhibitor Mini Tablets | Thermo Fisher Scientific | Cat # 88665 |
| Spheroid formation media | Irvine Scientific | Cat # 91130 |
| Polybrene | Santa Cruz | Cat # sc-134220 |
| Critical Commercial Assays | | |
| Subcellular Protein Fractionation Kit | Thermo Scientific | Cat # 78840 |
| Experimental Models: Cell Lines | | |
| Human pancreatic CAFs | https://www.ncbi.nlm.nih.gov/pmc/articles/PMC5283834/ | NA |
| NIH-3T3 fibroblasts | ATCC | Cat # CRL-1658™ |
| Panc-1 cells | ATCC | RRID:CVCL_0480 |
| hTERT-HPNE (hTERT) | ATCC https://www.ncbi.nlm.nih.gov/pubmed/17332339 | RRID:CVCL_C466 |
| hTERT-HPNE E6/E7/KRasG12D (KRAS) | ATCC https://www.ncbi.nlm.nih.gov/pubmed/17332339 | RRID:CVCL_C469 |
| Oligonucleotides | | |
| Non-targeting control siRNA | Life Technologies-Dharmacon | Cat # D-001810-01-05 |
| siRNA targeting ERK1 | Life Technologies-Dharmacon | Cat # L-003592-00-0005 |
| siRNA targeting ERK2 | Life Technologies-Dharmacon | Cat # L-003555-00-0005 |
| Software and Algorithms | | |
| ImageJ OrientationJ plugin software | http://bigwww.epfl.ch/demo/orientation/ | RRID:SCR_014796 |
| SMIA-CUKIE 2.1.0 | https://github.com/cukie/SMIA | RRID:SCR_014795 |
| MetaMorph 7.8.1.0 software | Molecular Devices | RRID:SCR_002368 |
| GraphPad Prism | https://www.graphpad.com/scientific-software/prism/ | RRID:SCR_002798 |

**Acknowledgements:**

The authors would like to dedicate this work to the memory of Dr. Patricia Keely whose work inspired this study. We thank Dr. E. Golemis for her input in revising and discussing this work, E. Ragan for proofing as well as S. Karamil, J. So and S. Eble for technical help. We thank Dr. R. Francescone for molecular biology consulting. Financial support was from the Commonwealth of Pennsylvania, Temple-FCCC’s Nodal Grant (EC, PIL, RM), NIH/NCI’s R01 CA113451 (EC), DOD W81XH-15-1-0170 (EC), funds from the Martin and Concetta Greenberg Pancreatic Cancer Institute (EC), and NIH/NCI’s CA06927 core grant supported facilities: Talbot Library, Biorepository, Translational, Histopathology, Biostatistics, Cell Imaging, Cell Culture, Instrument Shop and Glass Washing.

## Supplementary Figures

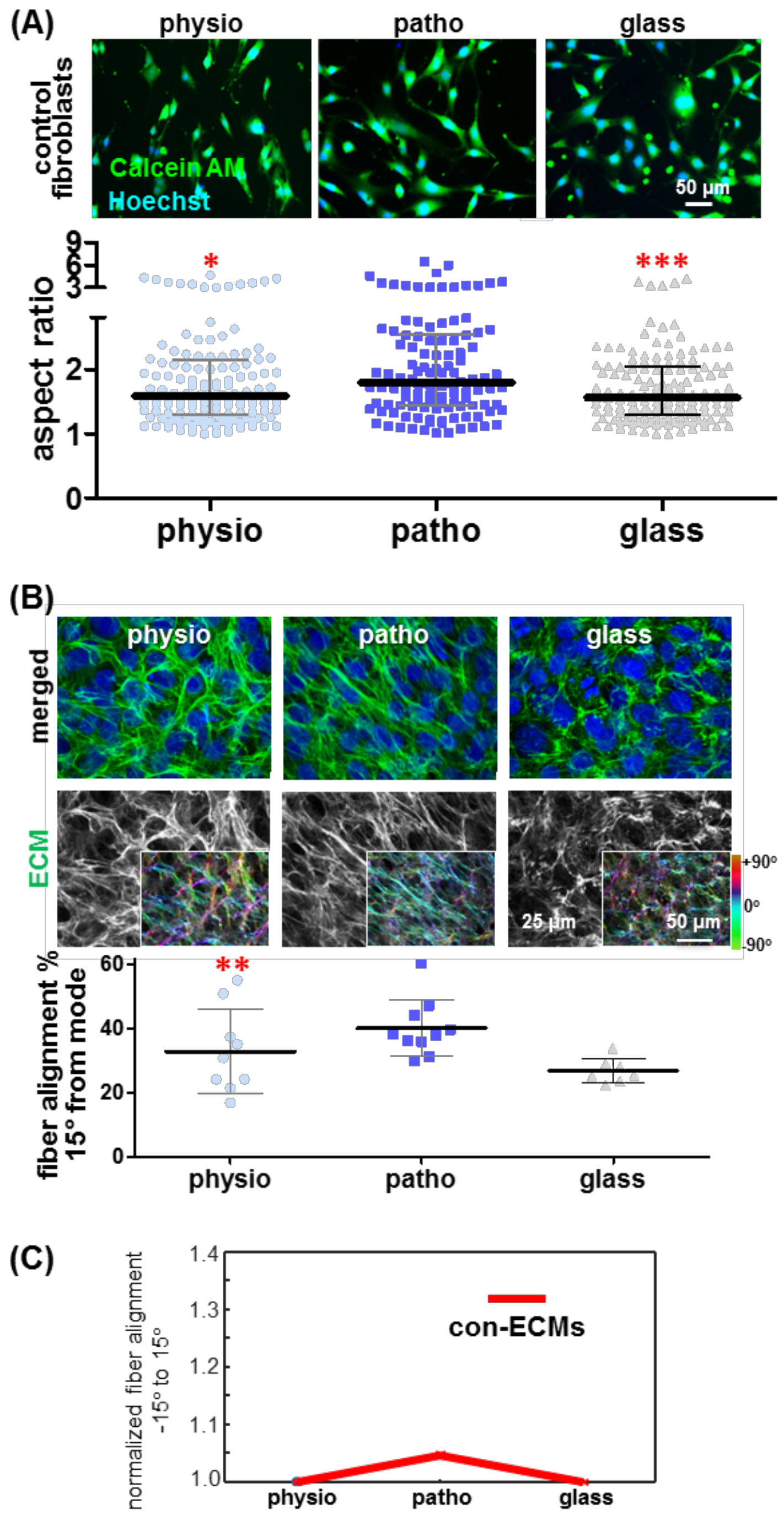

**Supplementary Figure. 1: Changes in substrate stiffness dictate a biphasic tendency of cell aspect ratio distribution in control fibroblast and corresponding ECM alignments dictated by underlying surface stiffness.** (**A**) Changes in substrate stiffness dictate a biphasic distribution of control fibroblast cell aspect ratios. Epi-fluorescent microscopy images depicting cell shape (CalceinAM; green) and nuclei (blue) of control fibroblasts cultured onto physio (~1.5 kPa), patho (~7 kPa) PA gels or glass coverslip. Graph (below) includes control aspect ratios (length/breadth) as calculated using MetaMorph software. Data is presented as median ± interquartile range. (**B**) (top) Reconstituted maximum projections corresponding to acquired confocal images of ECMs produced by control fibroblasts cultured onto physio- and patho-gels vs glass coverslips. Staining of fibronectin (green) and nuclei (blue) as well as monochromatic images are shown. Colors in bottom panel inserts depict angle distributions of ECMs, obtained using ‘OrientationJ’ plugin of Image-J software. Insert images were normalized using hue values for common/mode, cyan, and angle visualization as indicated by the gradient color bar on the right. Graph (below) shows percentages of fibers distributed at 15˚angles from the mode for the indicated experimental conditions. Note that the ECM fiber anisotropy is at peak levels when the matrix is produced on patho-gels. (**C**) Mathematical model rendition informed using the measured median aspect ratios of **A**, which were used to explain the observed biphasic distribution in ECM alignments, showing ratios of % angle distributions at 15 degrees from mode alignment, at the cell/matrix interphase.

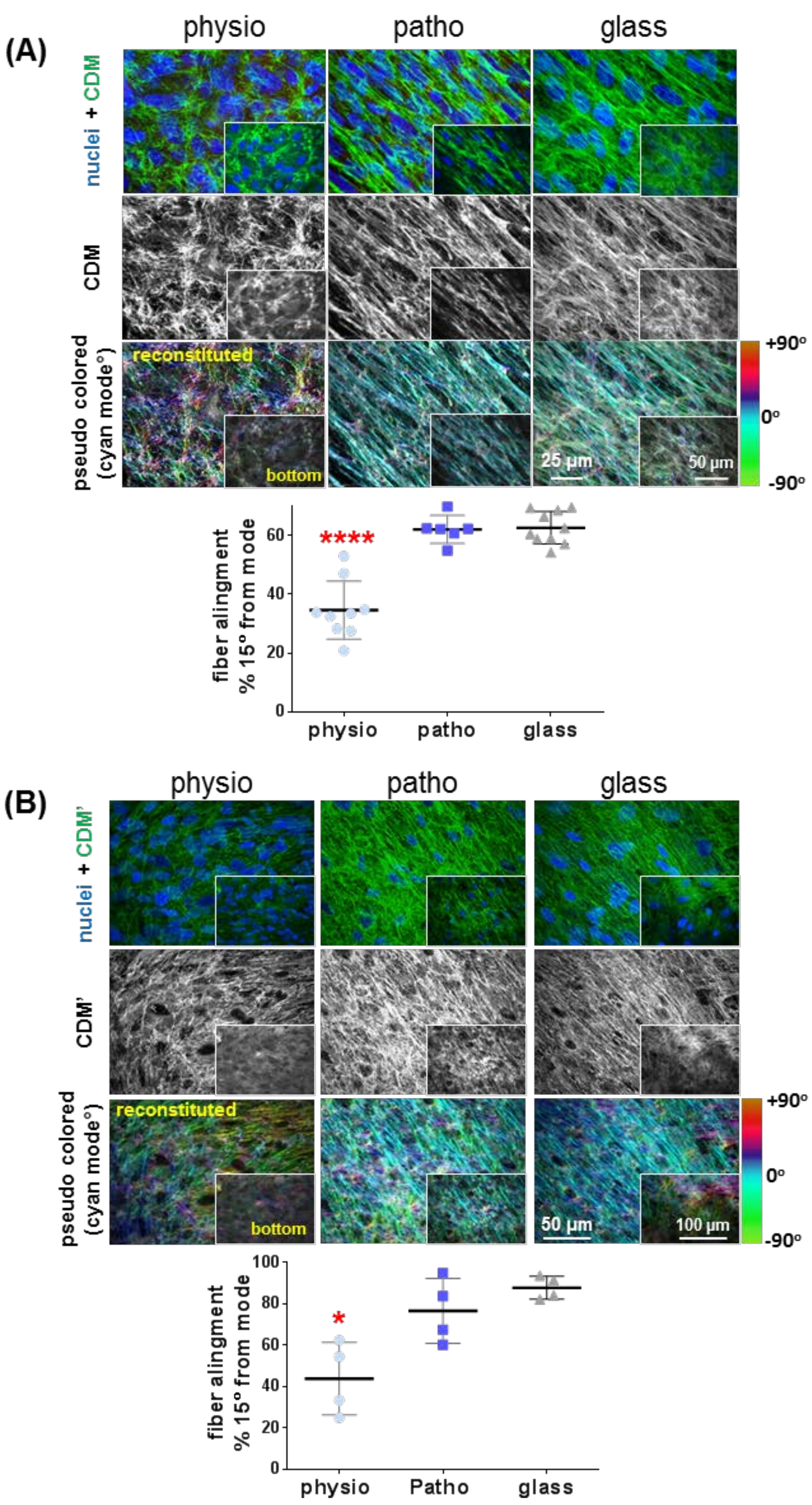

**Supplementary Figure 2: Patho and Physio-gels direct homogenous CDM production.** Reconstituted confocal microscopy images obtained from CDMs, produced by two independent CAF cells isolated from two different patients; **A** cells used through the rest of the study to generate CDMs and **B** additional PDAC patient-derived CAFs (**CDM'**). CAFs were plated onto the assorted substrates: 3-5 days on gels and 7-8 on glass coverslip. Staining of fibronectin (green) and nuclei (blue) and individual monochromatic images are shown (top row). Inserts represent only the bottom layers. Middle row includes the same CDM images in monochromatic depictions, omitting the nuclei channel. Colors in bottom panels depict angle distributions of CDMs, obtained using 'OrientationJ' plugin of Image-J software in which HUE was used to normalize colors to include, cyan, as the mode fibers for fiber angle distribution visualization, as indicated by the gradient color bar on the right. Graphs indicate percentage of CDM fibers oriented at 15 degrees from the mode measured angle. Asterisks denote: * $p<0.05$ and **** $p<0.001$.

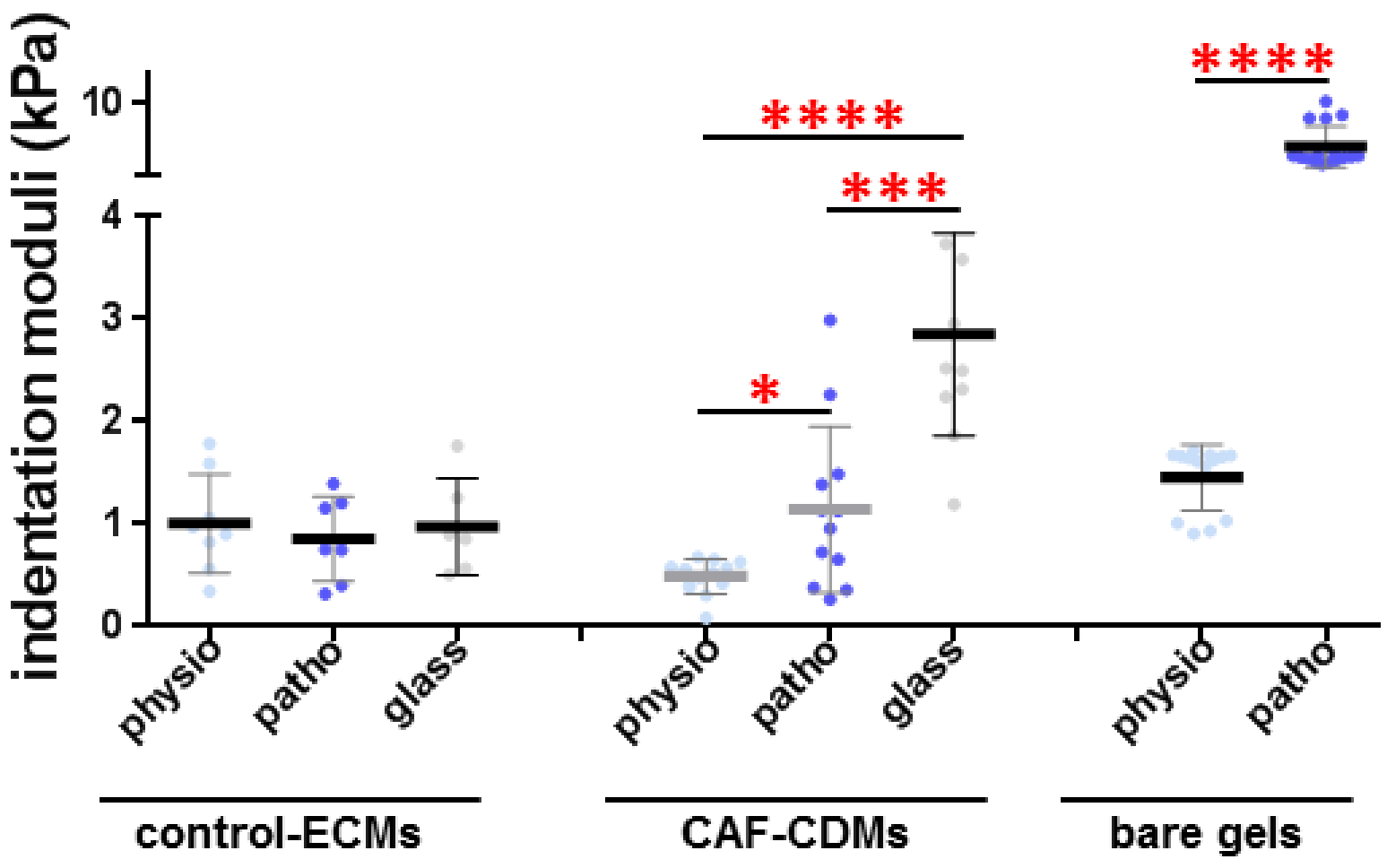

**Supplementary Figure 3: Indentation moduli of CDMs, but not control fibroblast produced ECMs, are linearly informed by the stiffness of their underlying substrates.** Indentation moduli of decellularized ECMs, produced by control fibroblasts or CAF-CDMs cultured onto the indicated substrates. Accompanying p-value table indicates the statistics obtained during indentation moduli measurements of decellularized CDMs generated by control fibroblasts or CAFs onto the assorted substrates using bare gels (physio and patho) as controls.

| substrate comparisons | significance | p values |
|---|---|---|
| physio vs patho | **** | <0.0001 |
| physio vs CAF-CDM physio | **** | <0.0001 |
| physio vs CAF-CDM patho | ns | 0.20 |
| physio vs CAF-CDM glass | **** | <0.0001 |
| physio vs control-ECM physio | * | 0.016 |
| physio vs control-ECM patho | ** | 0.0016 |
| physio vs control-ECM glass | * | 0.016 |
| patho vs CAF-CDM physio | **** | <0.0001 |
| patho vs CAF-CDM patho | **** | <0.0001 |
| patho vs CAF-CDM glass | **** | <0.0001 |
| patho vs control-ECM physio | **** | <0.0001 |
| patho vs control-ECM patho | **** | <0.0001 |
| patho vs control-ECM glass | **** | <0.0001 |
| CAF-CDM physio vs CAF-CDM patho | * | 0.0120 |
| CAF-CDM physio vs CAF-CDM glass | **** | <0.0001 |
| CAF-CDM physio vs control-ECM physio | ** | 0.003 |
| CAF-CDM physio vs control-ECM patho | * | 0.014 |
| CAF-CDM physio vs control-ECM glass | ** | 0.0049 |
| CAF-CDM patho vs CAF -CDM glass | *** | 0.0002 |
| CAF-CDM patho vs control-ECM physio | ns | p=0.67 |
| CAF-CDM patho vs control-ECM patho | ns | p=0.44 |
| CAF-CDM patho vs control-ECM glass | ns | p=0.65 |
| CAF-CDM glass vs control-ECM physio | *** | 0.0002 |
| CAF-CDM glass vs control-ECM physio | ns | p=0.22 |
| CAF-CDM glass vs control-ECM glass | *** | 0.0006 |
| control-ECM physio vs control-ECM patho | ns | p=0.43 |
| control-ECM physio vs control-ECM glass | ns | p=0.91 |
| control-ECM patho vs control-CDM glass | ns | p=0.48 |

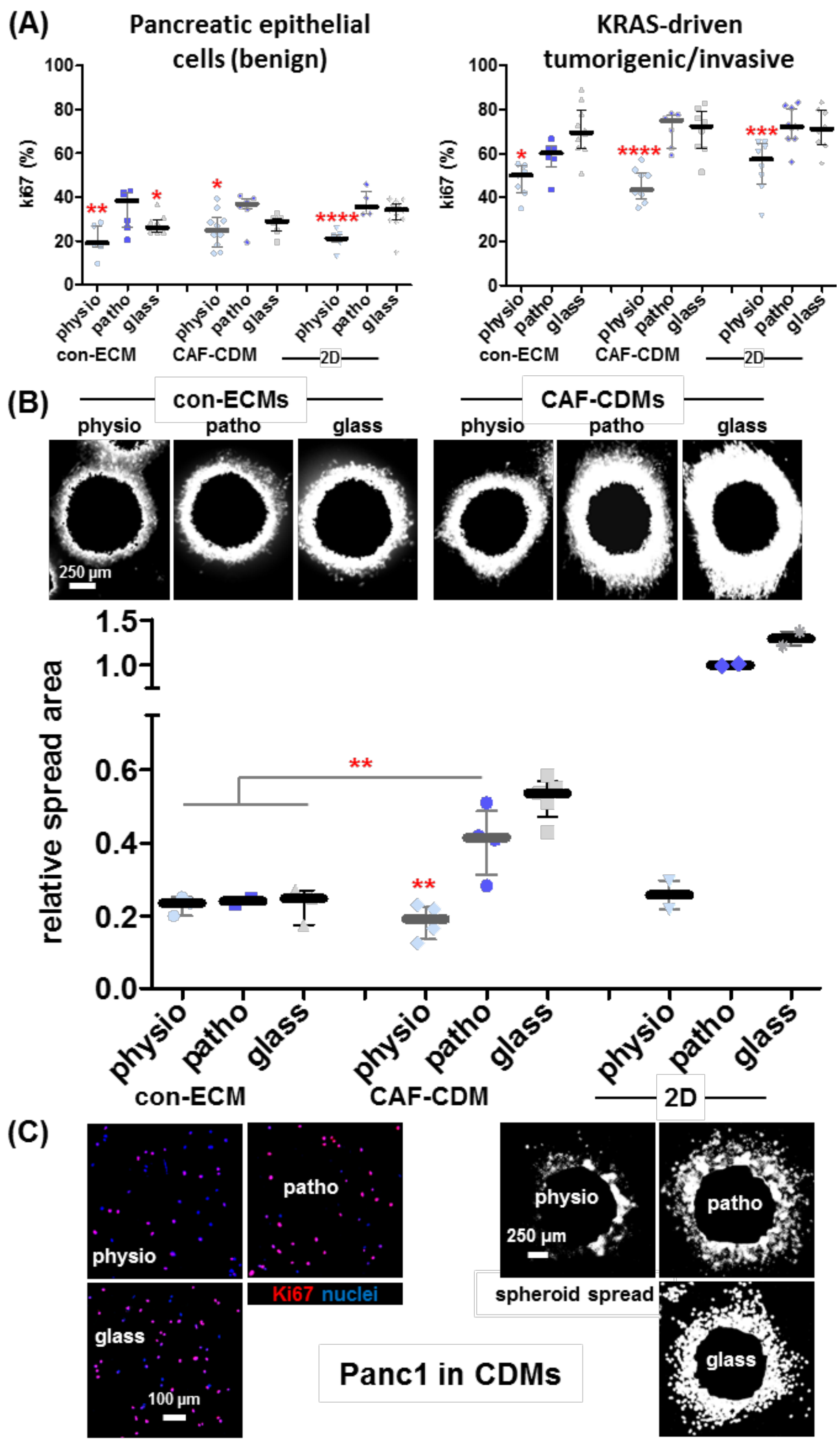

**Supplementary Figure 4: Tumor cell responses to 3D ECMs are restricted by isotropic CDMs.** (**A**) Syngeneic human pancreatic epithelial (benign) and KRAS-driven tumorigenic/invasive cancer cells were cultured within assorted fibroblaststic control ECM and CAF-CDMs or onto 2D substrates for 24 hrs. Ki67 levels were measured via indirect immunofluorescence using nuclei stain counts for normalization purposes and results (median ± interquartile range) were plotted. (**B**) RFP expressing tumorigenic/invasive cell spheroids, of even sizes, were allowed to spread on assorted control ECM and CAF-CDMs for 24 hrs. Images were acquired at times 0 and 24 hrs., and 95 percentile fluorescence intensities were used to measure relative (0 to 24) area spreads. Representative experimental images indicating the relative masked areas that were used to measure cell spread, which were plotted on the bottom graph, are shown. Asterisks denote: * $p<0.05$, ** $p<0.01$, *** $p<0.005$ and **** $p<0.001$. (**C**) Representative images of Ki67 incorporation (left; red) and nuclei (blue) as well as spheroid invasive spread (right) cultured in assorted CDMs as before, but this time with Panc1 PDAC cells.

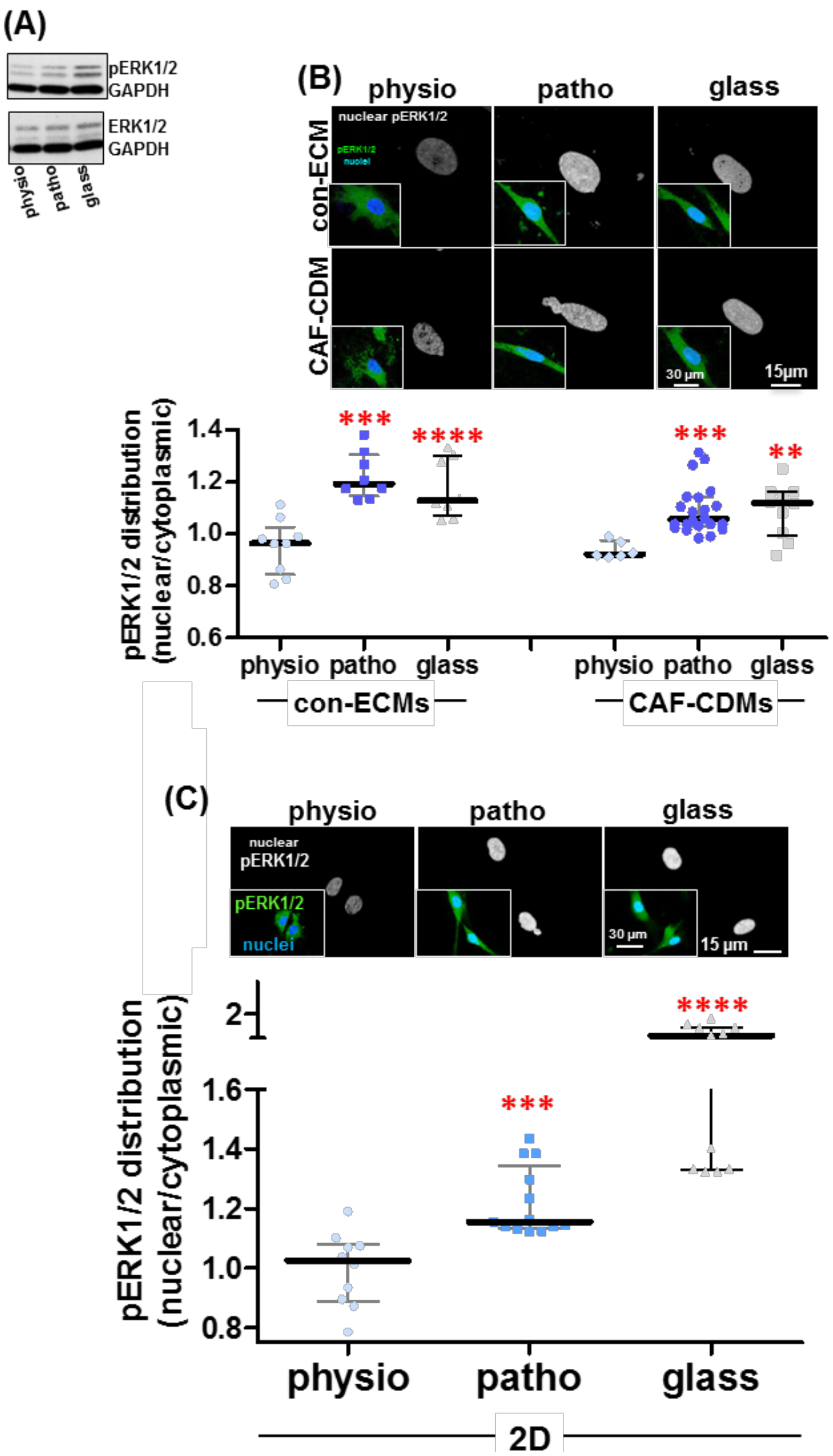

**Supplementary Figure 5: Distributions as opposed to levels of pERK1/2 are directed by CDMs.** (**A**) Western blot depicting the levels of pERK1/2 in cell lysates of KRAS-driven tumorigenic cultured on top of 2D assorted substrates. (**B**) Indirect immunofluorescence microscopy images of KRAS-driven tumorigenic cells cultured within the assorted ECMs. Monochromatic image of nuclear pERK1/2. Insert, on bottom left, shows how cells were double stained for total pERK1/2 (green) and nuclei (blue; used to select nuclei areas). Graph below depicts significant loss of nuclear pERK1/2 in both control-ECMs and CDM-ECMs that were generated on physio-gels (**C**). Indirect immunofluorescence and graph as in **B**, but cells were cultured onto 2D assorted substrates. Significance is shown with regards to corresponding physio-gel conditions.

**Supplemental Table**

| | MATRIX PROPERTIES | | | |
|---|---|---|---|---|
| | physio | patho | fold change | significance |
| gel stiffness (kPa) | 1.5 | 7.5 | 5 | **** |
| CDM stiffness (kPa) | 0.5 | 1.1 | ~2 | * |
| CDM alignment (% of fibers 15° from mode°) | 35 | 60 | ~2 | **** |
| | MATRIX INDUCED TUMOR CELL RESPONSES | | | |
| | CDM on physio | CDM on patho | (%) inhibition | significance |
| proliferation (%) | 45 | 71 | 37 | **** |
| migration (relative area spread) | 0.2 | 0.40 | 50 | ** |
| pERK1/2 localization (nuclear/cytoplasmic ratio) | 0.90 | 1.1 | 18 | *** |

Matrix properties included bare gels and CDMs made onto these and measuring stiffness, using atomic force microscopy as depicted in main text. Further, Matrix properties also included CDM fiber alignment measured by calculating the percentage of fibers aligned at 15 degrees from the mode angle. Fold changes were calculated dividing patho over physio gels or ECMs.
Matrix induced cell responses included Ki67 incorporation shown as percentage proliferation, spheroid spread areas shown as area spreads relative to spheroid core increases and portions of pERK1/2 nuclear over cytoplasmic intensity ratios. Percentage inhibition was calculated as proportion changes attained from CDMs made on pathological-gels compared to levels obtained imparted by CDMs produced on physio-gels. Significance is shown as * $p<0.05$, ** $p<0.01$, *** $p<0.005$ and **** $p<0.001$

**Movie:** CDMs produced onto glass coverslips showing alpha-smooth muscle actin, nuclei and CDM fibers. The movie shows each color in monochromatic mode plus overlay and plays a 0.5 micron slice per frame going from top to bottom. Note the differences in phenotypes observed at the beginning (top CDM layers) vs. end (interphase between bottom CDM production layer and the glass surface) of the movie. **https://drive.google.com/open?id=1DCPVw_sLPZ0dHpZ4rH7rBH8YQxRxMVJ_**